\documentclass[aps,prb,twocolumn,superscriptaddress,nofootinbib,citeautoscript,10pt]{revtex4-2}
\usepackage{amsmath,amssymb,bm,amsthm,amsfonts, makecell}
\usepackage{physics} 
\usepackage{mathrsfs,mathtools}

\usepackage{float}

\usepackage{xcolor}
\usepackage{svg}

\usepackage{graphicx}

\usepackage{bbold}

\usepackage[tight]{subfigure} 

\usepackage{color} 
\usepackage[papersize={8.5in,11in}]{geometry}
\usepackage[colorlinks=true]{hyperref}
\hypersetup{        
    unicode=false,          
    pdftoolbar=true,        
    pdfmenubar=true,        
    pdffitwindow=false,     
    pdfstartview={FitH},    
    pdfkeywords={keyword1} {key2} {key3}, 
    pdfnewwindow=true,      
    colorlinks=true,       
    linkcolor=magenta, 
    citecolor=blue,        
    filecolor=magenta,      
    urlcolor=blue           
} 

\usepackage{graphicx}
\usepackage{dcolumn}
\usepackage{color}
\usepackage{amssymb,amsmath,amsthm,amsfonts}
\usepackage{amsthm}
\usepackage{amsfonts}
\usepackage{tabularx,graphicx}
\usepackage{epstopdf}
\usepackage{latexsym}
\usepackage{colortbl}
\usepackage{psfrag}
\usepackage{bbm,bm,array,physics}
\usepackage{dsfont}
\usepackage{hyperref}

\begin{document}

\title{Superconducting diode effect in unconventional $p$-wave magnets}
\author{Igor de M. Froldi}

\affiliation{Instituto de F{\'i}sica, Universidade Federal de Goi{\'a}s, 74.690-900, Goi{\^a}nia-GO, Brazil}
\affiliation{Universit\'{e} Paris-Saclay, Institut de Physique Th\'eorique,  CEA, CNRS, F-91191 Gif-sur-Yvette, France}

\author{Hermann Freire}
\affiliation{Instituto de F{\'i}sica, Universidade Federal de Goi{\'a}s, 74.690-900, Goi{\^a}nia-GO, Brazil}

\begin{abstract} 
We investigate the emergence of superconducting phases, both with zero and finite Cooper-pair center-of-mass momenta, in recently proposed unconventional $p$-wave magnets. As a result, we find  that while these  magnetic phases are in principle compatible with a conventional pairing state at zero field, a Fulde-Ferrell phase can generally be promoted as the leading instability under the application of a finite magnetic field. Interestingly, by calculating the efficiency of the superconducting diode effect of this finite-momentum pairing state via a Ginzburg-Landau theory, we uncover that a high efficiency can be obtained in these systems for experimentally relevant spin splittings. Therefore, our prediction reveals that the experimental discovery of these materials represents a promising platform for the construction of superconducting logic circuits that may as a consequence find various technological applications.
\end{abstract}

\maketitle


\section{Introduction} 

Unconventional magnetism that features spin-split band structures with zero magnetization in the absence of magnetic field and spin-orbit coupling is a central research topic nowadays in condensed matter physics. One notable example is the recently discovered altermagnetism (AM) \cite{Smejkal2022_1,Smejkal2022_2,PhysRevX.12.040002,doi:10.7566/JPSJ.88.123702} that consists of a newly classified collinear magnetic state, which breaks the time-reversal symmetry $\mathcal{T}$ but does not exhibit net magnetization due to a compensating discrete rotational symmetry. As a result, it generates even-parity spin splittings ($d$, $g$, or $i$ wave) with many AM materials being subsequently discovered in the literature (see, e.g., Refs. \cite{krempaský_2023,Lee_PRL(2025),Osumi_PRB(2025),Ding_PRL(2024),Lu_2025,li_2024}). One of the motivations underlying this quest is that AM compounds are  expected to lead to important applications in the field of spintronics \cite{PhysRevLett.126.127701,PhysRevLett.129.137201,PhysRevX.12.011028,ch2025highly}. 
Shortly thereafter, inspired by this remarkable discovery, another unconventional magnetic state was proposed \cite{hellenes2024} for noncollinear coplanar magnetic moments that instead displays an odd-parity spin splitting \cite{hellenes2024,Minimal_models_2024,jungwirth2025,zk69-k6b2,85fd-dmy8,Maeda_2024,Cayao2025,CayaoPRB2025,Ezawa_2025,lin2025oddparity,zeng2025oddparityaltermagnetismspingroup,zhuang2025oddparity,AgterbergSC2025,fukaya_josephsoncurrent} ($p$, $f$, or $h$ wave). In the latter systems, the composite operation $\mathcal{T} \boldsymbol{\tau}$ (where $\boldsymbol{\tau}$ is a translation by half a unit cell of the lattice) is preserved but the inversion symmetry $\mathcal{I}$ is broken. In addition, the existence of such a composite symmetry $\mathcal{T}\boldsymbol{\tau}$ protects the odd-parity spin polarization, leading to a zero net magnetization. Some examples of compounds that might display unconventional $p$-wave magnetism (UPM) are, e.g., NiI$_2$ \cite{song2025electrical}, CeNiAsO \cite{hellenes2024,zk69-k6b2},  and Gd$_3$(Ru$_{1-x}$Rh$_x$)$_4$Al$_{12}$ \cite{Hirschberger_yamada2025}, while higher-order $h$-wave magnetism has recently been proposed in the context of some iron-based materials \cite{zk69-k6b2,85fd-dmy8}.

Furthermore, unconventional superconductivity (SC) arising as a result of proximity to magnetic phases is another topic of immense interest in the field, since a clear interconnection between these phases appears in many examples of strongly correlated quantum materials. Here, we focus specifically on the investigation of the impact of UPM on possible emergent unconventional SC phases. A similar analysis has been initiated in recent years regarding the effect of AM on new pairing states, with several interesting phenomena being predicted in a wide range of works, including, e.g., topological superconductivity \cite{Zhu2023,PhysRevLett.133.106601,PhysRevB.108.205410,PhysRevB.109.L201109}, pair-density waves (PDWs) \cite{Zhang_Nat_Comm_2024,Chakraborty2024,Knolle-arXiv(2024),Hong_PRB_2025,Hui_Hu_2025}, other unconventional SC phases \cite{Mazin2022notes,PhysRevB.108.224421,de_Carvalho_2024,Zhu2023,chakraborty2024constraintssuperconductingpairingaltermagnets,Mukasa_2025,wu2025intraunitcellsingletpairingmediated,heinsdorf2025}, the study of the superconducting diode effect (SDE) \cite{Ilic_Bergeret,Daido_2022,He_Nagaosa,Knolle-arXiv(2024),Yuan_2022,Scheurer-PRB(2024),chakraborty2024perfectsuperconductingdiodeeffect,froldi2025,shaffer2025theories_SDE}, etc. However, an analogous comprehensive investigation has not yet been performed in the context of UPM, except for a few recent works that studied, e.g., proximity effects of these new magnetic materials with superconductors \cite{Cayao2025,CayaoPRB2025,fukaya_josephsoncurrent,cm1l-1rsh,nagae2025majoranaflatbandsanomalous,Fukaya_JCMP}, the analysis of the coexistence of conventional superconductivity with UPM \cite{PhysRevB.111.L220403}, and the computation of transport properties associated with SC in these systems \cite{Kokkeler_2025}, among other examples (see, e.g., Ref. \cite{AgterbergSC2025}).

In this work, we derive a Ginzburg-Landau (GL) theory for the study of both zero- and finite-momentum superconductivity that appear in unconventional odd-parity magnets. Our analysis will concentrate on the investigation of nonreciprocal transport that takes place  specifically in emergent PDW phases. In this regard, we will study the SDE efficiency systematically as a function of the UPM spin splitting, temperature, and magnetic field. Consequently, we find, quite remarkably, that UPM materials are expected to yield highly efficient diode effects for experimentally relevant parameters, which could provide a promising pathway for advancing a new platform for energy-efficient electronic devices.

\section{Model and method} 

\begin{figure}[t]
    \centering
    \includegraphics[width=1\linewidth]{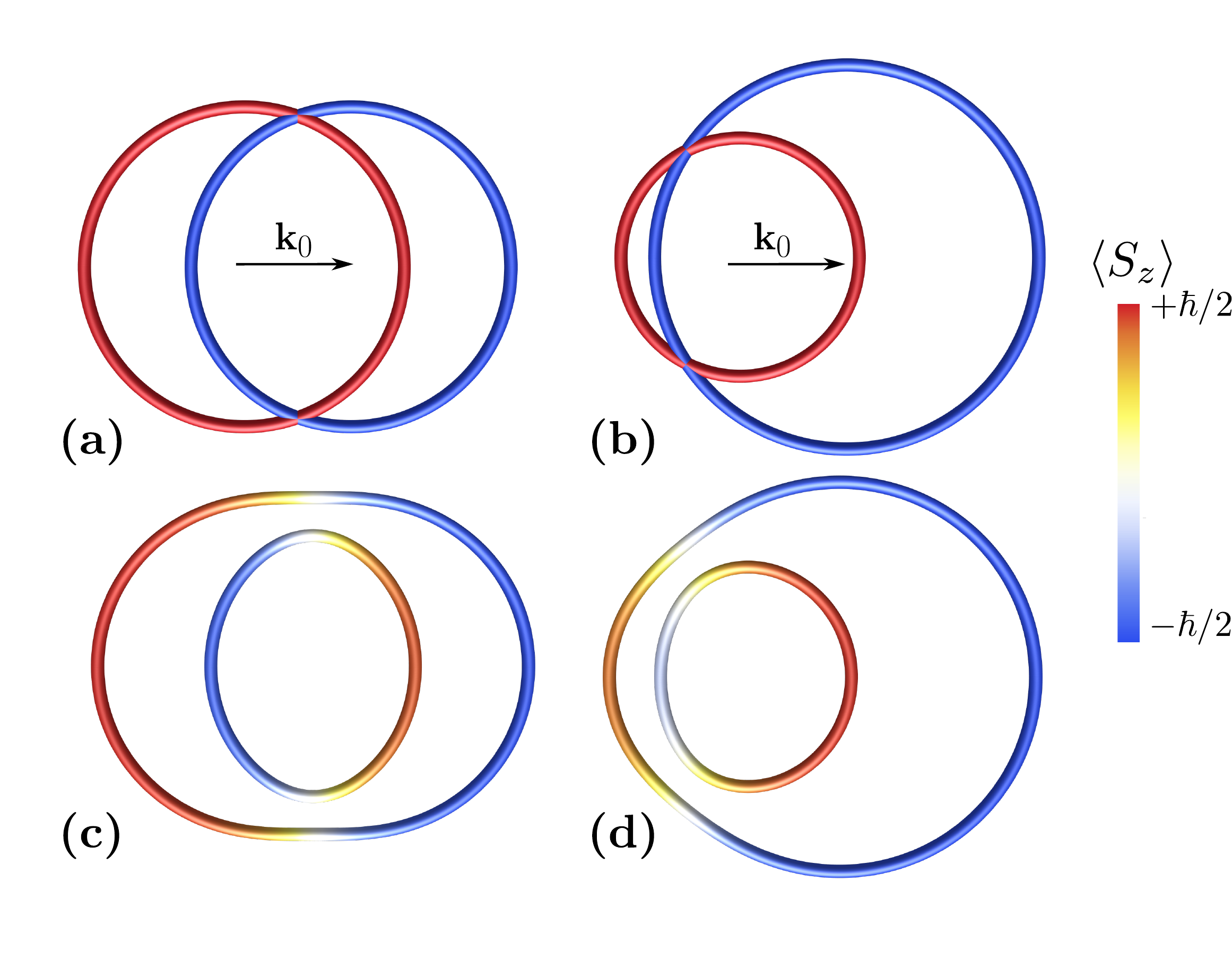}
     \caption{Schematic representation of the spin splitting of the Fermi surfaces near the $\Gamma$ point in a UPM, for different values of an out-of-plane magnetic field $\mathbf{B}=B_z \boldsymbol{\hat{z}}$ and $J_{sd}$ coupling: {(a)} $B_z=0$ and $J_{sd}=0$, {(b)} $B_z \neq 0$ and $J_{sd}=0$, {(c)} $B_{z}=0$ and $J_{sd} \neq 0$, and {(d)} $B_{z}\neq 0$ and $J_{sd} \neq 0$. The Fermi surfaces in (a) and (b) are displaced from the center of the zone for a finite $\lambda$, indicated by the wave vector $\mathbf{k}_0$.  Moreover, as $J_{sd}$ increases, the Fermi surfaces [i.e., (c) and (d)] become nodeless and acquire a nontrivial spin texture. The scale shown on the right denotes the spin projection $\langle S_z\rangle$ of all Fermi surfaces.} 
    \label{fig:1}
\end{figure}

We begin our analysis with a two-dimensional (2D) low-energy model that describes a $p$-wave itinerant magnet. To this end, we employ an effective model put forward in Ref. \cite{Minimal_models_2024} (see also Ref. \cite{hellenes2024}). 
The corresponding Hamiltonian is given by $H=\sum_{\mathbf{k}}c^{\dagger}_{\mathbf{k}}\mathcal{H}_{\text{UPM}}(\mathbf{k})c_{\mathbf{k}}$, where:
\begin{align}\label{HUPM}\nonumber
    \mathcal{H}_{\text{UPM}}(\mathbf{k}) &= -\left( 2t [\cos(k_x) + \cos(k_y)] + \tilde{\mu} \right)\sigma_0\otimes\tau_0\nonumber\\&+\left[\lambda_x\sin k_x+\lambda_y\sin k_y\right] \sigma_z\otimes\tau_0+J_{sd}\,\sigma_x\otimes\tau_z \nonumber\\&+ \mathbf{B}\cdot\boldsymbol{\sigma} \otimes \tau_0,
\end{align}
and $c_{\mathbf{k}}=(\psi_{\mathbf{k}A\uparrow},\psi_{\mathbf{k}B\uparrow},\psi_{\mathbf{k}A\downarrow},\psi_{\mathbf{k}B\downarrow})^{T}$, with $\psi_{\mathbf{k},i,s}$ ($\psi^{\dagger}_{\mathbf{k},i,s}$) being the annihilation (creation) operator of electrons with momentum $\mathbf{k}$, sublattice index $i=A,B$ and spin projection denoted by $s=\uparrow,\downarrow$, $t$ is the hopping amplitude for nearest neighbors, $\tilde{\mu}$ is the chemical potential, $\lambda_{x,y}$ generate the spin-helix structure, and $J_{sd}$ is the coupling between itinerant electrons and localized moments. Moreover, $\boldsymbol{\sigma}=(\sigma_{x},\sigma_{y},\sigma_{z})$ and $\tau_{x,y,z}$ are the Pauli matrices acting in the spin and sublattice spaces, respectively. For simplicity, we assume that $\lambda_x \equiv \lambda$ and $\lambda_y=0$, with the parameter $\lambda$ referring to the UPM spin splitting in the $k_x$ direction. The UPM form factor splits the bands at the Fermi level as shown schematically in Fig. \ref{fig:1}.

Unless otherwise specified, we will consider the regime in which the coupling between itinerant electrons and localized moments is much smaller than the UPM spin splitting ($J_{sd}\ll \lambda$). In this limit, the spin degrees of freedom decouple from the sublattice space. We will explain later in this work (and also in a more detailed way in Appendix \ref{Appendix_A}) the effect of a moderate $J_{sd}$ in the model. Therefore, in the former limit, the effective Hamiltonian becomes described near the $\Gamma$ point by:
\begin{align} \label{eq:AM}
    H_{0} &= \sum_{\mathbf{k},s,s'}\psi^{\dagger}_{\mathbf{k},s} [\mathbb{1}\xi_{\mathbf{k}} + \lambda\,\boldsymbol{g}_{\mathbf{k}} \cdot \boldsymbol{\sigma}+\mathbf{B} \cdot \boldsymbol{\sigma}]_{s,s'} \psi_{\mathbf{k},s'},
\end{align}
where $\psi^{\dagger}_{\mathbf{k},s}$ ($\psi_{\mathbf{k},s}$) creates (annihilates) an electron with momentum $\mathbf{k}$ and spin projection $s$, $\xi_{\mathbf{k}}={c_0\mathbf{k}^2} - \mu$, $\mathbb{1}$ is the identity matrix, $c_0=1/({2m})=t$, $\mu=\tilde{\mu}+4t$ is the (renormalized) chemical potential, and $m$ represents the effective mass. The strength of the spin splitting is parametrized by $\lambda$ with the form factor given by $ \boldsymbol{g}_{\mathbf{k}}$, which displays $p$-wave symmetry, i.e., $ \boldsymbol{g}_{\mathbf{k}}\sim k_x\boldsymbol{\hat{z}}$ in the vicinity of the $\Gamma$ point. Since the $p$-wave spin splitting satisfies $\boldsymbol{g}_{-\mathbf{k}} = -\boldsymbol{g}_{\mathbf{k}}$, the inversion symmetry $\mathcal{I}$ is broken and, therefore, this term is similar from this point of view to a Rashba spin-orbit coupling (we point out, though, that the former can be significantly larger than the latter due to its exchange interaction origin \cite{hellenes2024}). Moreover, we investigate here the effects of a uniform out-of-plane magnetic field $\mathbf{B}=B_z \boldsymbol{\hat{z}}$. Throughout this work, we will use units such that $\hbar=k_B=e=1$ in all expressions. Finally, we normalize the couplings at the noninteracting Fermi surface (FS) such that $\langle \abs{\boldsymbol{g}_{\mathbf{k}}}^2 \rangle_{\text{FS}} = 1$, where $\langle \cdot \cdot \cdot \rangle_{\text{FS}} = \int_{0}^{2 \pi} (\cdot \cdot \cdot) d \theta/2\pi$.

The pairing instabilities in this model can be analyzed by adding to Eq. \eqref{eq:AM} a standard BCS-type attractive interaction given by a general term:
\begin{align}\label{Hint}
     H_{\text{int}}=\frac{1}{2}\sum_{\substack{\mathbf{k},\mathbf{k'},\mathbf{q}\\s,s' }} V_{\mathbf{k},\mathbf{k}'} \psi^\dagger_{\mathbf{k}+\frac{\mathbf{q}}{2},s} \psi^\dagger_{-\mathbf{k}+\frac{\mathbf{q}}{2},s'} \psi_{-\mathbf{k}'+\frac{\mathbf{q}}{2},s'}  \psi_{\mathbf{k}'+\frac{\mathbf{q}}{2},s},
\end{align}
where $V_{\mathbf{k},\mathbf{k}'}<0$ near the Fermi surface, within a range defined by a characteristic cutoff $\epsilon_c$. For the moment, we will restrict our analysis to the possible formation of both zero- and finite-momentum spin-singlet $s$-wave superconductor [i.e., we will focus on the case in which $s=\uparrow$ and $s'=\downarrow$ in Eq. \eqref{Hint}]. This choice is motivated by the fact that this scenario, most commonly assumed for a phonon-mediated pairing mechanism, is a likely possibility of pair formation in the present model (see, e.g., the discussion in Ref. \cite{PhysRevB.111.L220403}). Consequently, we assume that $V_{\mathbf{k},\mathbf{k}'} =-V_s\,d_0(\mathbf{k})d_0(\mathbf{k'})$, where $V_s>0$ and $d_0(\mathbf{k})=1$. Therefore, we will concentrate our discussion in the main text on the investigation of such $s$-wave pairing states in the presence of UPM and leave the analysis of the possibility of the existence of long-range interactions that can promote spin-singlet $d_{xy}$-wave SC and spin-triplet $p$-wave pairing states to Appendices \ref{Appendix_A}, \ref{Appendix_B}, \ref{Appendix_C}, and \ref{Appendix_D}.

Next, we decouple the above interaction using mean-field theory in the singlet channel, by defining the order-parameter field self-consistently as follows:
\begin{align}
    \langle \psi_{-\mathbf{k}+\mathbf{q}/2,\downarrow} \psi_{\mathbf{k}+\mathbf{q}/2,\uparrow} \rangle \equiv -\sum_{\mathbf{k'}, \{\mathbf{Q}\}} \delta_{\mathbf{q},\mathbf{Q}} V^{-1}_{\mathbf{k},\mathbf{k}'} \,\Delta_{\mathbf{Q}}(\mathbf{k}'),
\end{align}
where we used the notation $\{\mathbf{Q}\}$ in the sum to make explicit the fact that the summation is discrete and takes place only on some specific values (to be determined shortly). Using the Nambu basis $\Psi_{\mathbf{k}} = ( \psi_{\mathbf{k}, \uparrow}, \psi_{\mathbf{k}, \downarrow}, \psi^{\dagger}_{-\mathbf{k},\uparrow}, \psi^{\dagger}_{-\mathbf{k},\downarrow})^T$, we write the complete Hamiltonian $H= H_0+H_{\text{int}}$ in the Bogoliubov-de Gennes (BdG) form:
\begin{align}\label{MicHamilt1}
    H &= \frac{1}{2}\sum_{\mathbf{k}, \mathbf{k}'} \Psi^{\dagger}_{\mathbf{k}'} 
            \begin{pmatrix}
                \mathcal{H}_0(\mathbf{k}) \delta_{\mathbf{k},\mathbf{k}'} & \Delta_{\mathbf{k},\mathbf{k}'}{(\mathbf{Q})} \\
                \left[ \Delta_{\mathbf{k}',\mathbf{k}}{(\mathbf{Q})} \right]^\dagger & - \left[ \mathcal{H}_0(\mathbf{k}) \right]^T \delta_{\mathbf{k}, \mathbf{k}'}
            \end{pmatrix} 
        \Psi_{\mathbf{k}} \nonumber\\& + \sum_{\mathbf{k},\mathbf{k}', \{ \mathbf{Q}\} }  V^{-1}_{\mathbf{k},\mathbf{k}'} |{\Delta_{\mathbf{Q}}(\mathbf{k}')}|^2,
\end{align} 
where a constant term has been dropped since it represents only a shift in total energy, and we have defined:
\begin{align}\label{FiedlDelta}
   \Delta_{\mathbf{k},\mathbf{k}'}{(\mathbf{Q})}= \sum_{  \{\mathbf{Q}\} } \Delta_{\mathbf{Q}} \left( \frac{\mathbf{k}+\mathbf{k}'}{2} \right) (-i \sigma_y) \delta_{\mathbf{k},\mathbf{k}'+\mathbf{Q}}.
\end{align}
The BdG Hamiltonian in Eq. \eqref{MicHamilt1} is quadratic in the fermionic fields. Hence, these degrees of freedom can be integrated out via a path-integral formalism, which will lead to the GL free energy (for details, see Appendix \ref{Appendix_B}). We  parametrize the order-parameter field as $\Delta^{(\mathbf{Q})}_s(\mathbf{k}) = |{\Delta_s^{(\mathbf{Q})}}| \,d_0(\mathbf{k})$, where $d_0(\mathbf{k})$ refers to the form factor already explained for the appropriate spin-singlet SC state, normalized over the Fermi surface. The resulting GL free energy up to fourth order in the order parameter yields:
\begin{align}\label{FreeEnergy}
    \mathcal{F}\left[ \Delta^{\dagger}, \Delta \right] &=  \sum_{\{\mathbf{Q}\}} \left(\alpha_{s}(\mathbf{Q},T) |{\Delta_s^{(\mathbf{Q})}}|^2 + \chi_{s} |{\Delta_s^{(\mathbf{Q})}}|^4 \right),
\end{align}
where we neglected, without loss of generality, the free energy in the normal state. The coefficient $\alpha_{s} (\mathbf{Q},T)$ is given by a polarization bubble in the appropriate pairing channel and describes the phase transition from the normal metal phase to the superconducting states (with either zero or finite Cooper-pair momentum). By defining the effective coupling $\mathbf{g}^{\text{eff}}_{\mathbf{k}} = \lambda\boldsymbol{g}_{\mathbf{k}}+\mathbf{B}$ and using the digamma function $\psi^{(0)}(x)$ to define $\Xi(x)=\Re[\psi^{(0)}\left( \frac{1}{2} + ix \right)]-\psi^{(0)}\left(\frac{1}{2} \right)$, this second-order coefficient becomes:
\begin{align}\nonumber
    \alpha_s(\mathbf{q},T) =& N_F\ln\left( \frac{T}{T^{(s)}_{c,0}} \right) + \frac{N_F}{4} \sum_{\gamma=\pm} \int_{FS} \frac{d\theta}{2\pi} d_0^2(\mathbf{k})\\\nonumber
    \Bigg[  &  \Xi \left( \frac{\gamma g_1 + \frac{\mathbf{q}}{2} \cdot \mathbf{v_F}}{2 \pi T}  \right) \left( 1 + \widehat{\mathbf{g}}^{\text{eff}}_{\mathbf{k}+\frac{\mathbf{q}}{2}} \cdot \widehat{\mathbf{g}}^{\text{eff}}_{-\mathbf{k}+\frac{\mathbf{q}}{2}} \right)\\\label{SingletCoef}
    +  & \Xi \left( \frac{\gamma g_2 + \frac{\mathbf{q}}{2} \cdot \mathbf{v_F}}{2 \pi T}  \right)  \left( 1 - \widehat{\mathbf{g}}^{\text{eff}}_{\mathbf{k}+\frac{\mathbf{q}}{2}} \cdot \widehat{\mathbf{g}}^{\text{eff}}_{-\mathbf{k}+\frac{\mathbf{q}}{2}} \right)  \Bigg],
\end{align}
where {$\widehat{\mathbf{g}}^{\text{eff}}_{\mathbf{k}}\equiv{\mathbf{g}}^{\text{eff}}_{\mathbf{k}}/|{\mathbf{g}}^{\text{eff}}_{\mathbf{k}}|$} is the unit vector along the effective coupling vector, while $g_{1,2} =( |{\mathbf{g}^{\text{eff}}_{\mathbf{k}+\mathbf{q}/2}}| \pm |{\mathbf{g}^{\text{eff}}_{-\mathbf{k}+\mathbf{q}/2}}| )/2$ denotes, respectively, the mean energy and the energy difference due to the effective coupling at different points on the Fermi surface, and $\mathbf{v_F} = k_F(\cos\theta\,\boldsymbol{\hat{x}} + \sin\theta\,\boldsymbol{\hat{y}})  /m$ stands for the Fermi velocity. The coupling strength $V_s$ in the spin-singlet channel has been absorbed in the definition of the critical temperature $T^{(s)}_{c,0}$ in the absence of spin-splitting and magnetic fields ({i.e.}, for $\lambda=\abs{\mathbf{B}}=0$) defined as
$T^{(s)}_{c,0} = ({2e^{\gamma} \epsilon_c}/{\pi}) e^{-1/(N_F V_s)}$,
where $\gamma= 0.57721...$ is the Euler-Mascheroni constant and $N_F$ is the noninteracting density of states at the Fermi level. Henceforth, all temperatures and couplings will be measured w.r.t. $T^{(s)}_{c,0}$. The coefficient $\chi_s$ in Eq. \eqref{FreeEnergy} is derived for our microscopic model from a four-legged Feynman diagram, which is calculated for the corresponding minimizing wave vector $\mathbf{Q}$. Since the latter result yields a cumbersome expression, we leave its detailed explanation for Appendix \ref{Appendix_C}.  

We are now ready to define the set denoted by $\{ \mathbf{Q} \}$, e.g., in Eq. \eqref{MicHamilt1}. They can be extracted by minimizing the coefficient $\alpha_{s}{(\mathbf{q},T)}$ with respect to an arbitrary momentum $\mathbf{q}$ and temperature $T$, thus obtaining a temperature-dependent wave vector $\mathbf{Q}(T)$, similar to that found for the case of PDW formation in electronic liquid-crystal phases \cite{Soto_2014}. We then find the corresponding critical temperature (denoted by $T^{(s)}_{c}$) for the transition between the normal state and the spin-singlet SC phases by solving the linearized gap equation, {i.e.,} $\alpha_{s}[\mathbf{Q}(T^{(s)}_{c}), T^{(s)}_{c}]=0$. Interestingly, we obtain that, at zero magnetic field, a UPM splitting does not generate any finite-momentum pairing for spin-singlet states, which contrasts with the scenario regarding the emergence of superconductivity in altermagnets \cite{Zhang_Nat_Comm_2024,Chakraborty2024,Knolle-arXiv(2024),Hong_PRB_2025,Hui_Hu_2025,froldi2025}.
\begin{figure}[t]
    \centering
    \includegraphics[width=0.75\linewidth]{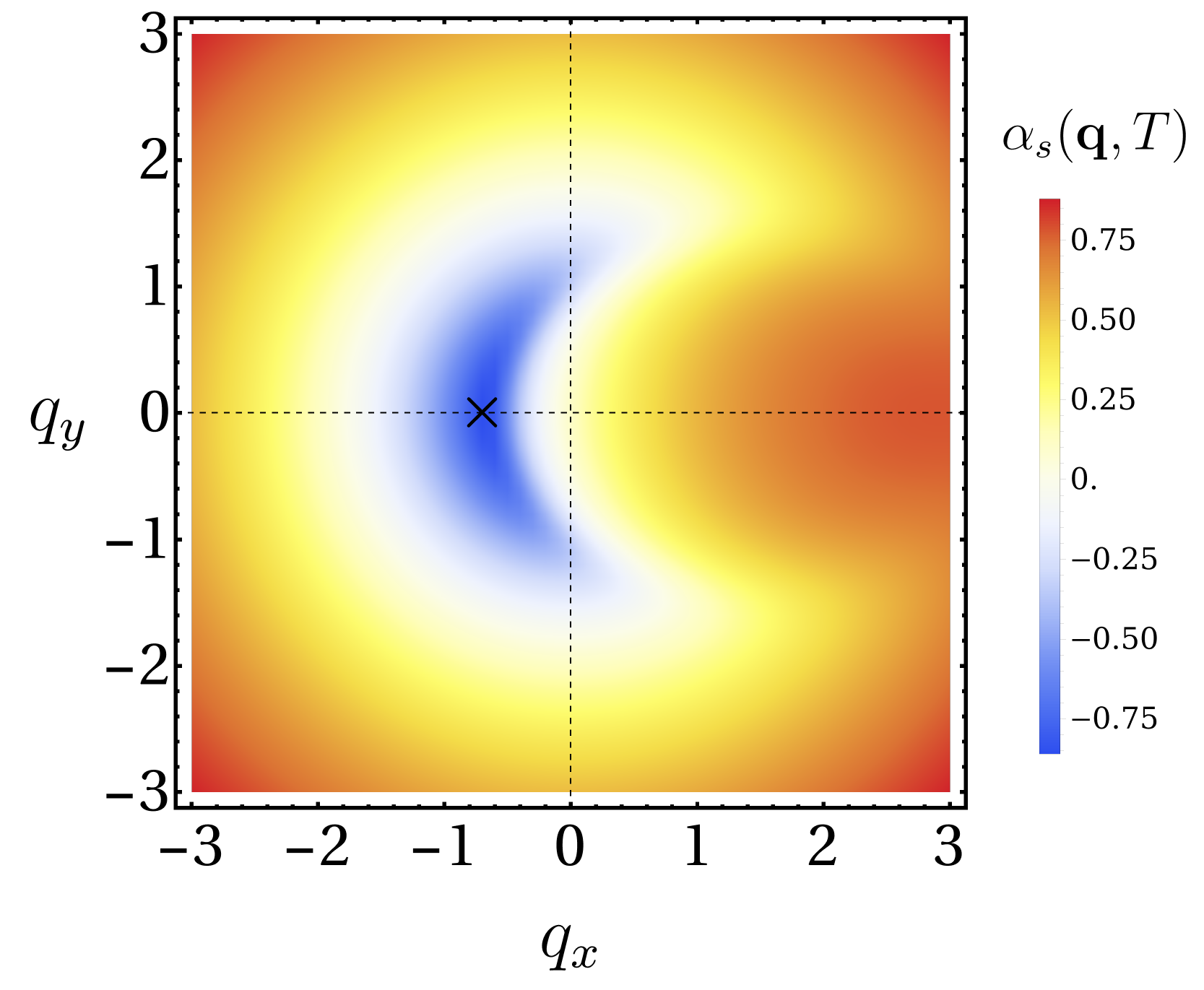}
    \caption{Second-order GL coefficient [Eq. \eqref{FreeEnergy}] for the $s$-wave FF state in the present model. Here, we choose $\lambda/T^{(s)}_{c,0}=1$, $B_{z}/T^{(s)}_{c,0}=1$, and $T=0.1T^{(s)}_{c,0}$. The symbol ``$\boldsymbol{\times}$" marks the global minimum of the GL free energy, which yields $\mathbf{Q}\approx -0.7 \mathbf{\hat{x}}$ in this case. The ``half-moon" pattern of this plot stems from both time-reversal and inversion symmetry breakings.}
    \label{Density}
\end{figure}

As we switch on the magnetic field, only one stable minimum referring to a pairing phase with modulation $\mathbf{Q}$ is found. The direction of $\mathbf{Q}$ turns out to be temperature independent (for details, see Appendix \ref{Appendix_B}), which points along one of the antinodal directions (see Fig. \ref{Density}) -- again, in sharp contrast to the case of altermagnets \cite{froldi2025,Knolle-arXiv(2024),chakraborty2024perfectsuperconductingdiodeeffect}. This spatial modulation implies that the only finite-momentum superconducting phase that emerges in this case is a Fulde-Ferrell (FF) state \cite{Fulde_Ferrell,Agterberg_2020}, which breaks spontaneously both time-reversal and inversion symmetries. The corresponding expression that describes the FF order parameter is then given by $\Delta^{\text{FF}}_s(\mathbf{r})=\Delta^{(\mathbf{Q})}_se^{i\mathbf{Q}\cdot\mathbf{r}}$. The minimization of the free energy, Eq. \eqref{FreeEnergy}, with respect to the order-parameter field for the FF modulation is then given by:
\begin{align}
    \mathcal{F}_{\text{FF}}^{\text{min}} = - \frac{\alpha_s(\mathbf{Q},T)}{4 \chi_s(\mathbf{Q},T)},
\end{align}
where the stability conditions inside the pairing state are given by $\alpha_s[\mathbf{Q}(T),T)] <0$ and $\chi_s[\mathbf{Q}(T),T]>0$.

\begin{figure*}[t]
        \centering
        \includegraphics[width=0.96\linewidth]{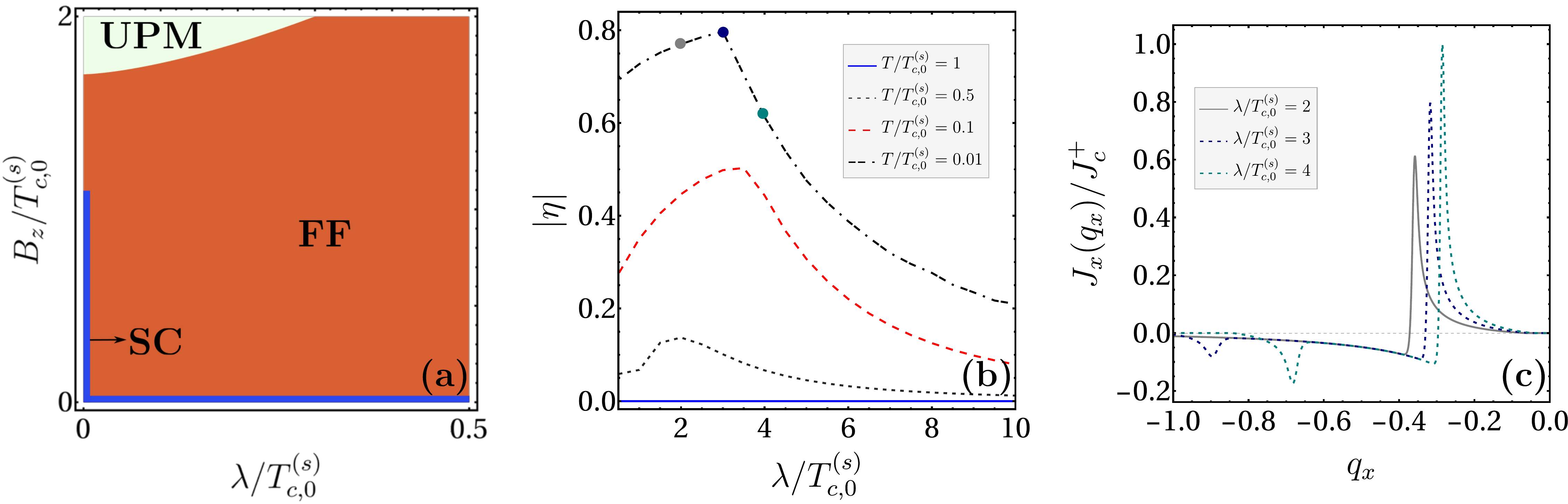}
        \caption{{(a)} Schematic pairing phase diagram of $B_z$ vs $\lambda$ in the present model  for $J_{sd}\ll \lambda$. The white region denotes the UPM phase, the red region stands for the spin-singlet $s$-wave FF phase and the blue region corresponds to the spin-singlet $s$-wave zero-momentum SC phase. {(b)} Efficiency $\eta $ of the SDE [Eq. \eqref{efficiency}] for the FF phase with modulation along the direction of the minimizing wave vector $\mathbf{Q}$ as a function of the UPM spin splitting $\lambda$ for a finite magnetic field such that $ B_z/T^{(s)}_{c,0} =1$. We also show the behavior of $\eta$ in the model for different temperatures $T$. {(c)} Supercurrent as a function of the momentum $q_x$ for different values of the spin splitting around the maximum of the efficiency, indicated by three circles shown in (b), at  $T=0.01T^{(s)}_{c,0}$.}
        \label{PhaseDiagram}
    \end{figure*}

\section{Phase diagram} 

We now turn our attention to the determination of the phase diagram associated with our effective model [Eq. \eqref{eq:AM}], which is displayed in Fig. \ref{PhaseDiagram}(a). For $\mathbf{B}=0$, since the UPM Hamiltonian displays explicitly only inversion symmetry breaking, a spin-singlet pairing state remains unaffected w.r.t. the increase of $\lambda$. 
As we switch on the magnetic field $\mathbf{B}$ along the out-of-plane ($\boldsymbol{\hat{z}}$) direction, and for $\lambda=0$, the zero-momentum spin-singlet pairing state is suppressed, but only above a finite critical field. For nonzero $\lambda$ and $\mathbf{B}$, the situation changes dramatically. Even for a small magnetic field, the increase in $\lambda$ leads to the formation of an FF state, indicating that UPM is a favorable platform for the emergence of finite-momentum superconductivity in the spin-singlet channel. As $\mathcal{T}$-symmetry breaking terms are detrimental to spin-singlet pairs, there is a maximum $B_{P}$, which suppresses completely both zero and finite-momentum pairing in this case (this is the so-called Pauli limiting field). Note that $B_P$ is enhanced as we increase the UPM splitting $\lambda$, as indicated by the line separating the FF and the UPM phases in Fig. \ref{PhaseDiagram}(a). This behavior is similar with the properties discussed by works on superconductivity in noncentrosymmetric materials, where an inversion-asymmetric Rashba spin-orbit coupling (SOC) also has the effect of increasing the Pauli field (see, e.g., Ref. \cite{Frigeri_2004}).

\section{Diode effect}

\begin{figure}[t]
    \centering
    \includegraphics[width=0.85\linewidth]{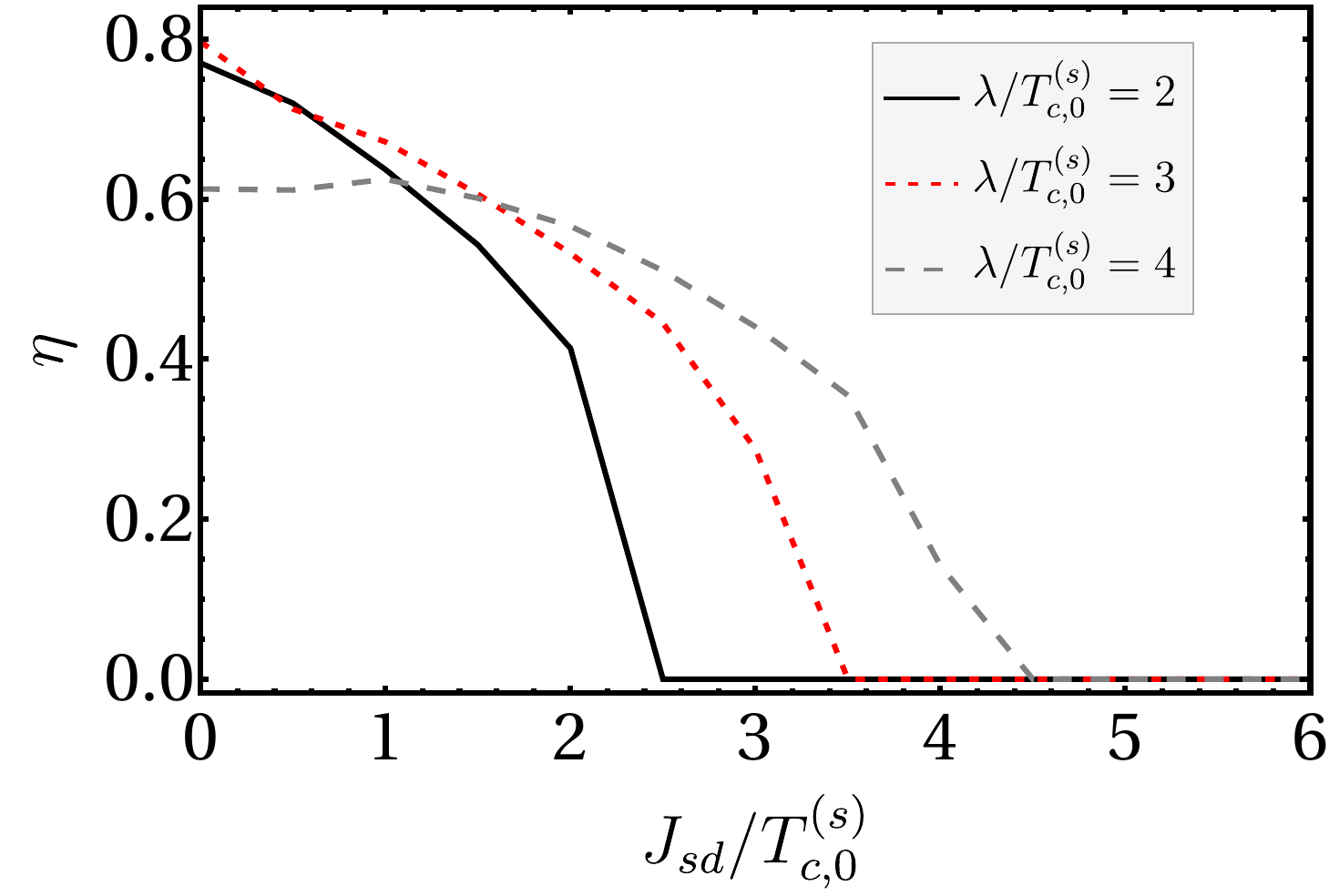}
    \caption{SDE efficiencies as a function of the $J_{sd}$ coupling in Eq. \eqref{HUPM} associated with different values of the UPM spin splitting $\lambda$ for the spin-singlet $s$-wave FF phase with modulation along the direction of the minimizing wave vector $\mathbf{Q}$, a finite magnetic field $B_z/T_{c,0}^{(s)}=1$, and temperature $T/T_{c,0}^{(s)}=0.01$.}
    \label{EffJsd}
\end{figure}

Since the UPM effective model clearly exhibits (even for a small $\lambda$ and $\mathbf{B}$) a spin-singlet FF pairing phase with both time-reversal and inversion symmetry breakings, nonreciprocal phenomena are naturally expected to emerge. The supercurrent associated with such a phase displayed by our effective model can be calculated via minimal coupling of the superconducting wave vector to an external vector potential $\mathbf{A}$ using the relation $\mathbf{J}(\mathbf{q})= - \lim_{\mathbf{A} \rightarrow 0} \nabla_{\mathbf{A}} \mathcal{F}(\mathbf{q}-2 \mathbf{A})$. Therefore, we obtain:
    \begin{align}
    \mathbf{J}(\mathbf{q}) = -\frac{1}{2 \chi_s} \alpha_{s}(\mathbf{q},T) \nabla_{\mathbf{q}} \alpha_{s}(\mathbf{q},T).
    \end{align}
For nonzero $\lambda$ and $\mathbf{B}$, the supercurrent is not only asymmetric with respect to $\mathbf{q}$ but can also be skewed, leading to different maximum and minimum values. These features displayed by our model naturally lead to nonreciprocal properties such as the superconducting diode effect (SDE), which refers to a phenomenon where the critical currents parallel and antiparallel to the Cooper-pair momentum turn out to be unequal \cite{levitov_5259947,Edelstein_1996,ando2020observation,He_Nagaosa,Daido_2022,shaffer2025theories_SDE}. By writing $\mathbf{J}(\mathbf{q}) = J(\mathbf{q}) \boldsymbol{\hat{n}}$, we can analyze this supercurrent in fixed directions $\boldsymbol{\hat{n}}$ with respect to the direction of the minimizing wave vector $\mathbf{Q}$. Defining $J_c^{+}$ and $J_c^{-}$ as, respectively, the maximum and minimum values of the supercurrent, we can evaluate the asymmetry of the supercurrent in the model by computing the SDE efficiency $\eta$ given by:
\begin{align}\label{efficiency}
    \eta = \frac{J^{+}_c-J^{-}_c}{J^{+}_c + J^{-}_c}.
\end{align}
In Fig. \ref{PhaseDiagram}(b), we plot this quantity for different temperatures as a function of the UPM splitting $\lambda$ for a finite magnetic field (here, we assume that $\boldsymbol{\hat{n}}=\boldsymbol{\hat{x}}$, which points along the direction of the modulation $\mathbf{Q}$ of the FF phase -- this yields the maximized SDE efficiency). Consequently, we observe that the SDE efficiency has a peak for low temperatures at an optimal splitting $\lambda_{{opt}}$, which can reach almost $80\%$ of efficiency for some conveniently chosen physical parameters. We also point out that the SDE efficiency clearly exhibits a monotonically increasing behavior with decreasing temperature. As can be seen in Fig. \ref{PhaseDiagram}(c), where the supercurrents are plotted at $T=0.01T^{(s)}_{c,0}$ for UPM splittings $\lambda$ with either smaller or larger values than this optimal point, the supercurrent starts to develop a local minimum and the variation of $\lambda$ makes this a global minimum responsible for the decrease in the parameter $\eta$. We also note that the efficiency goes to zero when $\lambda = 0$, indicating that UPM is crucial to obtain a finite SDE in our model. Lastly, we point out that as the $J_{sd}$ coupling in Eq. \eqref{HUPM} increases, we observe that the SDE is clearly diminished in UPM systems, as shown in Fig. \ref{EffJsd}. This arises due to the fact that the $J_{sd}$ has a tendency to suppress the $s$-wave FF phase in the corresponding effective model and, consequently, it drives a transition from such a pairing phase to the normal state of the UPM for large enough $J_{sd}$. In addition, we also note that as the spin splitting $\lambda$  increases, the SDE efficiency remains significant even for larger values of the $J_{sd}$ coupling (for more information, see also Appendix \ref{Appendix_F}).

Recently, density-functional-theory (DFT) calculations have predicted UPM, e.g., in the heavy-fermion compound CeNiAsO \cite{hellenes2024}, and also in some iron-based superconductors (see, e.g., LaFeAsO \cite{AgterbergSC2025}). We point out that in the latter compounds, in the presence of SOC, a $p$-wave spin splitting is obtained along the $k_x$ and $k_y$ directions, while in the $k_z$ direction it exhibits an $h$-wave character. The UPM splitting $\lambda$ in the iron-based superconductor was predicted to be relatively small ($\lambda \sim 10$ meV), while in the heavy-fermion case it was found to be one order of magnitude larger ($\lambda \sim 100$ meV). Our calculation indicates that, for larger ratios $(\lambda/T^{(s)}_{c,0})$, the diode effect is expected to have a higher efficiency with the application of the magnetic field at low temperatures. Therefore, specifically for the candidate material LaFeAsO, the reported spin splitting falls within the range $\lambda \sim 1 - 10$ meV, which fits into the window of highest SDE efficiencies for the case of $T^{(s)}_{c,0} \sim 30$ K [see again Fig. \ref{PhaseDiagram}(b)]. 

Before arriving at the final conclusions, we make a brief comment, for the sake of completeness, on possible emergent $d_{xy}$-wave spin-singlet SC and $p$-wave pairing states in the presence of UPM within our model. 
For the case of $d_{xy}$-wave SC, we have verified that the corresponding pairing phase diagram does not change qualitatively compared to the $s$-wave pairing scenario: for $B_z=0$, we find a zero-momentum SC phase for any $\lambda$, and, as soon as both $\lambda$ and $\mathbf{B}$ are finite, we obtain that the leading instability is toward a spin-singlet $d_{xy}$-wave FF phase, whose SDE efficiency yields  smaller values than the $s$-wave FF case (for more details, see Appendix \ref{Appendix_E}). 

Regarding possible spin-triplet pairings in the model, we have investigated the instabilities associated with $p$-wave symmetry for both mixed spin (i.e., chiral-type), described by the form factor vector $\boldsymbol{d}({\mathbf{k}})=(k_x+i k_y)\boldsymbol{\hat{z}}$, and equal spin (i.e., helical-type), described by the order-parameter $\boldsymbol{d}({\mathbf{k}})=i(k_x\boldsymbol{\hat{y}}+k_y\boldsymbol{\hat{x}})$ \cite{Bernevig2013}. For $\mathbf{B}=0$, the chiral (mixed spin) $p$-wave state does not generate any finite-momentum superconductivity, while the helical (equal spin-triplet) is capable of generating PDW at zero fields with two minimizing wave vectors $\pm \mathbf{Q}$. The latter corresponds to a spatial modulation given by ${\Delta}^{\text{LO}}_t(\mathbf{r})={\Delta}^{(\mathbf{Q})}_te^{i\mathbf{Q}\cdot\mathbf{r}}+{\Delta}^{(-\mathbf{Q})}_t e^{-i\mathbf{Q}\cdot\mathbf{r}}$ (also called a Larkin-Ovchinnikov (LO) phase \cite{Agterberg_2020,osti_4653415}), therefore preserving both time-reversal and inversion symmetries; for this reason, no SDE efficiency is obtained for this case. The analysis of energetics within the GL theory confirms that this helical spin-triplet pairing phase indeed stabilizes the LO state. Upon the application of a finite magnetic field, the LO phase is strongly suppressed and gives way to the corresponding zero-momentum SC state, while the chiral spin-triplet $p$-wave state develops a finite Cooper-pair momentum with an FF modulation, whose SDE efficiency is quantitatively similar to the FF spin-singlet $s$-wave state counterpart. All technical details on the above analyses and also the investigation of the effect of increasing $J_{sd}$ in the phase diagram and the SDE efficiency of the triplet pairing are explained in Appendices \ref{Appendix_D}, \ref{Appendix_E}, and \ref{Appendix_F}.

\section{Summary and outlook} 

In the present work, we have systematically studied the possible emergence of superconductivity with either zero or finite Cooper-pair center-of-mass momentum in recently proposed unconventional $p$-wave magnets from a GL theory perspective. Consequently, we have shown that, while for zero magnetic field a SC phase is obtained, an FF phase appears as the leading instability for a finite $\mathbf{B}$. By computing the efficiency of the SDE of the finite-momentum pairing phase via the GL analysis, we have found that a high efficiency of the diode effect is expected in these systems for experimentally relevant UPM spin splittings. Therefore, our present results indicate that the experimental discovery of these materials represents a promising platform for the design of energy-efficient logic circuits that may find interesting technological applications \cite{shaffer2025theories_SDE}. 

Since it has recently been demonstrated from a theoretical perspective that iron-based superconductors with coplanar magnetic order are likely odd-parity magnets \cite{zk69-k6b2,85fd-dmy8,AgterbergSC2025}, these compounds may provide a concrete materialization of odd-parity magnetism coexisting with superconductivity. In the specific case of the material LaFeAsO doped with phosphorus, coplanar magnetic order was experimentally shown to coexist with superconductivity within a particular region of the phase diagram \cite{stadel_1906390}. Therefore, this compound represents to date the best candidate material for observing all the features investigated here, including a highly efficient SDE under the application of a magnetic field. For this reason, we hope that our present work can stimulate further experimental activities on the remarkable properties of such unconventional odd-parity magnetic compounds.

\emph{Note added---} Recently, {we became aware of two related works} discussing the possibility of emergent superconducting phases in systems displaying UPM  (see Refs. \cite{khodas2026superconductivitypwavemagnets,pal2026emergent}).

\section*{Acknowledgments} H.F. acknowledges funding from the Conselho Nacional de Desenvolvimento Cient\'{i}fico e Tecnol\'{o}gico (CNPq) under Grant No. 404274/2023-4 {and No. 305575/2025-2}. We also acknowledge the support of the INCT project Advanced  Quantum Materials, involving the Brazilian agencies CNPq (Proc. 408766/2024-7), FAPESP (Proc. 2025/27091-3), and CAPES.

\appendix

\section{Additional details regarding the methodology}
\label{Appendix_A}

\begin{figure*}[t]
    \centering
    \includegraphics[width=1\linewidth]{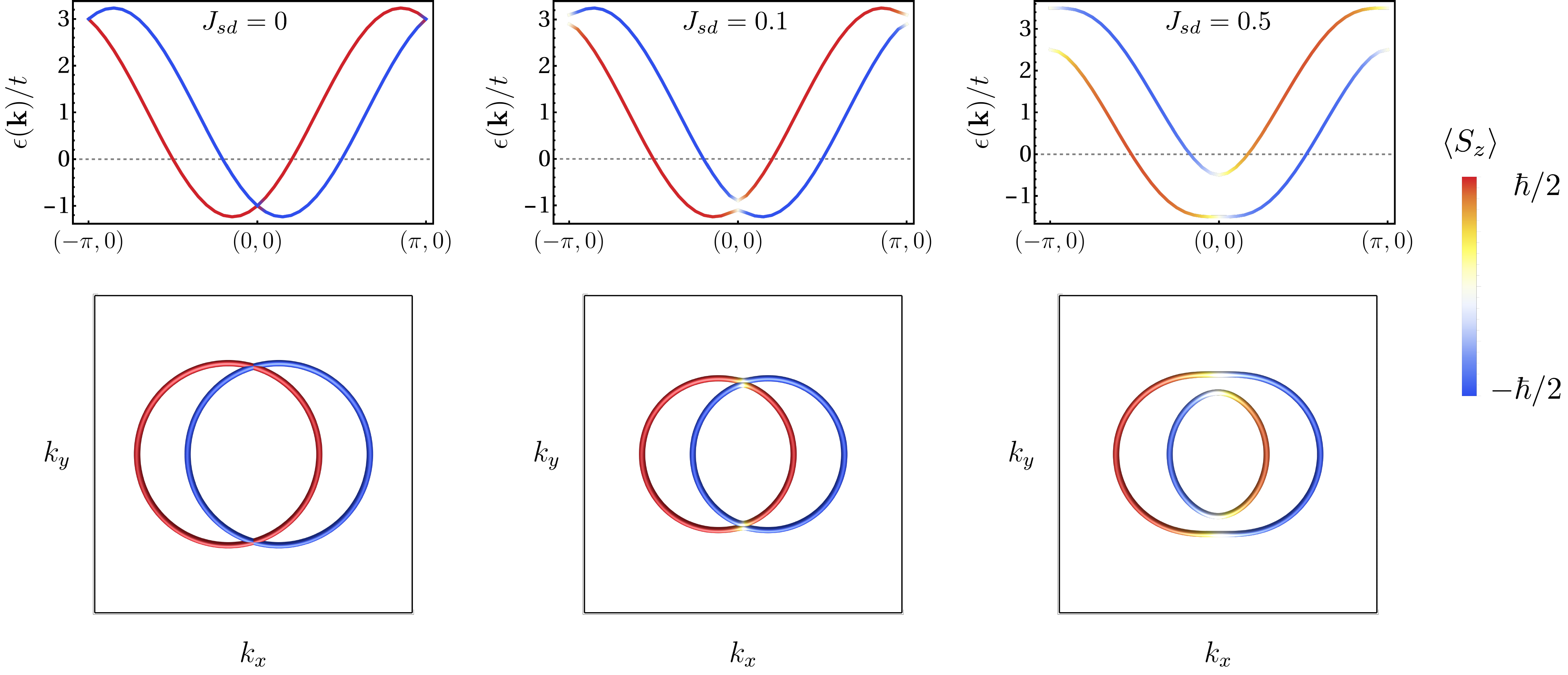}
    \caption{Band structure associated with the Hamiltonian in Eq. \eqref{HUPM} along a high-symmetry path for different values of the $J_{sd}$ coupling (in units of $t$) and the corresponding Fermi surfaces in the vicinity of the $\Gamma$ point (for $\mathbf{B}=0$). The colors of the energy bands and of the Fermi surfaces indicate the spin texture, given by the scale of the spin polarization $\langle S_z\rangle$ shown on the right. For the above plots, we have set $\lambda=t$ and $\tilde{\mu}=-3t$.}
    \label{BandsJsd}
\end{figure*}

As explained in the main text, the effective Hamiltonian proposed for unconventional $p$-wave magnets \cite{Minimal_models_2024,hellenes2024} is given by Eq. \eqref{HUPM}. For simplicity, we assume that $\lambda_x \equiv \lambda$ and $\lambda_y=0$, with the parameter $\lambda$ referring to the UPM spin-splitting in the $k_x$ direction. First, we will consider the regime in which the coupling between itinerant electrons and localized moments is much smaller than the UPM spin splitting (i.e., $J_{sd}\ll \lambda$) and leave the discussion on how the results change for a moderate $J_{sd}$ to Appendix \ref{Appendix_F}. In the former limit, the spin degrees of freedom decouple from the sublattice space. As a result, the effective model for the unconventional $p$-wave magnet (UPM) given in the main text becomes described near the $\Gamma$ point by the  Hamiltonian given by:
\begin{align}
    H_0 = \sum_{\mathbf{k}, s, s'} \psi^\dagger_{\mathbf{k},s} \left[ \mathbb{1}\xi_\mathbf{k} + \lambda \boldsymbol{g}_{\mathbf{k}} \cdot \boldsymbol{\sigma} + \mathbf{B} \cdot\boldsymbol{\sigma} \right]_{s,s'}  \psi_{\mathbf{k},s'} \equiv \sum_{\mathbf{k}}\mathcal{H}_0(\mathbf{k}),
\end{align}
where $\psi^{\dagger}_{\mathbf{k},s}$ ($\psi_{\mathbf{k},s}$) creates (annihilates) an electron with momentum $\mathbf{k}$ and spin projection $s$ and the energy dispersion is $\xi_{\mathbf{k}} = \mathbf{k}^2/(2m) - \mu$, where $\mu=\tilde{\mu}+4t$. This Hamiltonian describes an example of a UPM (we write $\boldsymbol{g}_{\mathbf{k}} = k_x \boldsymbol{\hat{z}}$ in the vicinity of the $\Gamma$ point) parametrized by the spin-splitting strength $\lambda$. In Fig. \ref{BandsJsd}, we show that there is little difference in the Fermi surfaces around the $\Gamma$ point between these two models in the limit of $J_{sd}\ll \lambda$ (however, we point out that for moderate $J_{sd}$ these differences become significant). We also define the effective coupling $\mathbf{g}^{\text{eff}}_{\mathbf{k}}=\lambda \boldsymbol{g}_\mathbf{k} +\mathbf{B}$. The superconducting instabilities are then studied by adding a BCS-type attractive interaction given by:
\begin{align}
    H_{\text{int}} = \frac{1}{2} \sum_{\mathbf{k}, s,s'} V_{\mathbf{k},\mathbf{k}'} \psi^\dagger_{\mathbf{k}+\mathbf{q}/2,s} \psi^\dagger_{-\mathbf{k}+\mathbf{q}/2,s'} \psi_{-\mathbf{k}'+\mathbf{q}/2,s'}  \psi_{\mathbf{k}'+\mathbf{q}/2,s},
\end{align}
where $V_{\mathbf{k},\mathbf{k}'}<0$, and we have taken the volume of the system to be equal to unity. We  decouple the above interaction in the pairing channel by defining the pairing function $\widehat{\mathscr{F}}_{\mathbf{Q}}(\mathbf{k}) $:
\begin{align}
    \langle \psi_{-\mathbf{k}+\mathbf{q}/2,s'} \psi_{\mathbf{k}+\mathbf{q}/2,s} \rangle \equiv \sum_{\{\mathbf{Q}\}} \delta_{\mathbf{q},\mathbf{Q}} \left[ \widehat{\mathscr{F}}_{\mathbf{Q}}(\mathbf{k}) i \sigma_y  \right]_{s,s'},
\end{align}
and the sum in $\mathbf{Q}$ is discrete over some particular values (to be determined later). We define the order-parameter field $\widehat{\Delta}_{\mathbf{Q}}(\mathbf{k})$ as:
\begin{align}
    \widehat{\Delta}_{\mathbf{Q}}(\mathbf{k}) \equiv- \sum_{\mathbf{k}'} V_{\mathbf{k},\mathbf{k}'} \widehat{\mathscr{F}}_{\mathbf{Q}}(\mathbf{k}') \Rightarrow \widehat{\mathscr{F}}_{\mathbf{Q}}(\mathbf{k}) = - \sum_{\mathbf{k}'} V^{-1}_{\mathbf{k}',\mathbf{k}} \widehat{\Delta}_{\mathbf{Q}}(\mathbf{k}').
\end{align}
Using the Nambu basis $\Psi_{\mathbf{k}} = ( \psi_{\mathbf{k}, \uparrow} \quad \psi_{\mathbf{k}, \downarrow} \quad \psi^{\dagger}_{-\mathbf{k},\uparrow} \quad \psi^{\dagger}_{-\mathbf{k},\downarrow})^T$, we write the full Hamiltonian $H= H_0+H_{\text{int}}$ as:
\begin{align}\nonumber
    H = &\frac{1}{2}\sum_{\mathbf{k}, \mathbf{k}'} \Psi^{\dagger}_{\mathbf{k}'} 
            \begin{pmatrix}
                \mathcal{H}_0(\mathbf{k}) \delta_{\mathbf{k},\mathbf{k}'} & \Delta(\mathbf{k},\mathbf{k}',\mathbf{\mathbf{Q}}) \\
                \left[ \Delta(\mathbf{k}',\mathbf{k},\mathbf{\mathbf{Q}}) \right]^\dagger & - \left[ \mathcal{H}_0(-\mathbf{k}) \right]^T \delta_{\mathbf{k}, \mathbf{k}'}
            \end{pmatrix} 
        \Psi_{\mathbf{k}} \\\label{MicHamilt1-Appendix}
        &+ \frac{1}{2} \sum_{\mathbf{k},\{ \mathbf{Q}\}} \tr \left( \left[ \widehat{\mathscr{F}}_{\mathbf{Q}}(\mathbf{k}) \right]^\dagger \widehat{\Delta}_{\mathbf{Q}}(\mathbf{k}) \right),
\end{align} 
where:
\begin{align}\label{FiedlDelta}
    \Delta(\mathbf{k},\mathbf{k}',\mathbf{Q}) = \sum_{\{  \mathbf{Q} \}} \widehat{\Delta}_{\mathbf{Q}} \left( \frac{\mathbf{k}+\mathbf{k}'}{2} \right)(-i \sigma_y) \delta_{\mathbf{k},\mathbf{k'}+\mathbf{Q}}.
\end{align}

The microscopic action is then obtained by $\mathcal{S}[\Bar{\psi},\psi,\Bar{\Delta},\Delta] = \int_{0}^{\beta} d \tau \left( \frac{1}{2} \Psi^\dagger \partial_\tau \Psi + H \right)$, where $\beta=1/T$ is the inverse temperature. In Matsubara frequency space, we get for the microscopic action:
\begin{widetext}
\begin{align}\nonumber
    \mathcal{S} =&  \frac{1}{2} T\sum_{\mathbf{k}, \mathbf{k}',i\omega_n} \Psi_{\mathbf{k}'}^{\dagger}
            \begin{pmatrix}
                \left[ -i\omega_n+\mathcal{H}_0(\mathbf{k}) \right] \delta_{\mathbf{k},\mathbf{k}'} & \Delta(\mathbf{k},\mathbf{k}',\mathbf{\mathbf{Q}}) \\
                \left[ \Delta(\mathbf{k}',\mathbf{k},\mathbf{\mathbf{Q}}) \right]^\dagger & \left[-i\omega_n - \left[ \mathcal{H}_0(-\mathbf{k}) \right]^T \right] \delta_{\mathbf{k}, \mathbf{k}'}
            \end{pmatrix} 
        \Psi_{\mathbf{k}} + \frac{\beta}{2} \sum_{\mathbf{k},\{ \mathbf{Q}\}} \tr \left( \left[ \widehat{\mathscr{F}}_{\mathbf{Q}}(\mathbf{k}) \right]^\dagger \widehat{\Delta}_{\mathbf{Q}}(\mathbf{k}) \right)\\\label{MicAction1}
        \equiv& \frac{1}{2} T\sum_{\mathbf{k}, \mathbf{k}',i\omega_n} \Psi_{\mathbf{k}'}^{\dagger} \left( \left[ \widehat{\mathcal{G}}_0(\mathbf{k},i \omega_n) \right]^{-1} \delta_{\mathbf{k},\mathbf{k}'} + \widehat{\mathcal{M}}(\mathbf{k},\mathbf{k}',\mathbf{Q}) \right) 
        \Psi_{\mathbf{k}} + \frac{\beta}{2} \sum_{\mathbf{k},\{ \mathbf{Q}\}} \tr \left( \left[ \widehat{\mathscr{F}}_{\mathbf{Q}}(\mathbf{k}) \right]^\dagger \widehat{\Delta}_{\mathbf{Q}}(\mathbf{k}) \right),
\end{align} 
\end{widetext}
where a constant term was dropped since it represents
only a shift in total energy. Also, we defined the inverse of the free propagator $ \widehat{\mathcal{G}}_0(\mathbf{k},i \omega_n) $ (with $\omega_n=(2n+1)\pi T, n \in \mathbb{N}$) and the order-parameter matrix is given by $\widehat{\mathcal{M}}(\mathbf{k},\mathbf{k}',\mathbf{Q})$. Written in terms of the particle and hole propagators, the inverse of the free propagator becomes:
\begin{widetext}
\begin{align}
    \left[ \widehat{\mathcal{G}}_0(\mathbf{k},i \omega_n) \right]^{-1}\equiv
            \begin{pmatrix}
                \left[ -i\omega_n+\mathcal{H}_0(\mathbf{k}) \right] \delta_{\mathbf{k},\mathbf{k}'} & 0 \\
                0 & \left[-i\omega_n - \left[ \mathcal{H}_0(-\mathbf{k}) \right]^T \right] \delta_{\mathbf{k}, \mathbf{k}'}
            \end{pmatrix} =
            \begin{pmatrix}
                \left[ G_0^{(p)} \left( \mathbf{k},  i\omega_n \right) \right]^{-1} & 0 \\
                0 & \left[ G_0^{(h)} \left( \mathbf{k},  i\omega_n \right) \right]^{-1}
            \end{pmatrix},
\end{align}
\end{widetext}
where:
\begin{align}
    G_0^{(p)} \left( \mathbf{k},  i\omega_n \right) =& \frac{1}{ \left( -i\omega_n + \xi_{\mathbf{k}} \right) \sigma_0 + \mathbf{g}^{\text{eff}}_{\mathbf{k}} \cdot \boldsymbol{\sigma} },\\
    G_0^{(h)} \left( \mathbf{k}, i\omega_n \right) =& \frac{1}{ \left( -i\omega_n - \xi_{\mathbf{-k}} \right) \sigma_0 +  \mathbf{g}^{\text{eff}}_{\mathbf{-k}} \cdot \sigma_y \boldsymbol{\sigma} \sigma_y }
\end{align}
can be further simplified to:
\begin{widetext}
\begin{align}\label{ParticlePropagator}
    G^{(p)}_0 \left(\mathbf{k}, i \omega_n \right) =& G^{(+)} \left( \mathbf{k}, i\omega_n \right) \sigma_0 + G^{(-)} \left( \mathbf{k}, i\omega_n \right) \mathbf{\widehat{g}}^{\text{eff}}_{\mathbf{k}} \cdot \boldsymbol{\sigma},\\\label{HolePropagator}
    G^{(h)}_0 \left(\mathbf{k}, i \omega_n \right) =& - G^{(+)} \left( -\mathbf{k}, -i\omega_n \right) \sigma_0 + G^{(-)} \left( -\mathbf{k}, -i\omega_n \right) \mathbf{\widehat{g}}^{\text{eff}}_{\mathbf{-k}} \cdot \sigma_y \boldsymbol{\sigma} \sigma_y,
\end{align}
(where {$\widehat{\mathbf{g}}^{\text{eff}}_{\mathbf{k}} \equiv{\mathbf{g}}^{\text{eff}}_{\mathbf{k}}/|{\mathbf{g}}^{\text{eff}}_{\mathbf{k}}|$}) by defining:
\begin{align}
    G^{(+)} \left( \mathbf{k}, i\omega_n \right) =& \frac{1}{2} \left( \frac{1}{-i \omega_n + \xi_{\mathbf{k}}+ \abs{\mathbf{g}^{\text{eff}}_{\mathbf{k}}}} + \frac{1}{-i \omega_n + \xi_{\mathbf{k}} - \abs{\mathbf{g}^{\text{eff}}_{\mathbf{k}}}} \right),\\
    G^{(-)} \left( \mathbf{k}, i\omega_n \right) =& \frac{1}{2} \left( \frac{1}{-i \omega_n + \xi_{\mathbf{k}} + \abs{\mathbf{g}^{\text{eff}}_{\mathbf{k}}}} - \frac{1}{-i \omega_n + \xi_{\mathbf{k}} - \abs{\mathbf{g}^{\text{eff}}_{\mathbf{k}}}} \right).
\end{align}
\end{widetext}
We now work out the last term in Eq. \eqref{MicAction1}. In order to present a complete picture of possible SC phases emerging from a UPM, we decompose the pairing order-parameter matrix as:
\begin{align}
    \widehat{\Delta}^{(\mathbf{q})}(\mathbf{k}) = \Delta_s^{(\mathbf{q})}(\mathbf{k}) \sigma_0 + \mathbf{\Delta}_t^{(\mathbf{q})}(\mathbf{k})\cdot \boldsymbol{\sigma},
\end{align}
where $\Delta_s^{(\mathbf{q})}(\mathbf{k})$ and $\boldsymbol{\Delta}_t^{(\mathbf{q})}(\mathbf{k})$ refer to the singlet and triplet pairing amplitudes, respectively. We take $\Delta_s^{(\mathbf{q})}(\mathbf{k})$ and $\boldsymbol{\Delta}_t^{(\mathbf{q})}(\mathbf{k})$ to obey:
\begin{align}\label{eq_s14}
    \sum_{\mathbf{k}'} V_{\mathbf{k},\mathbf{k}'} \Delta_s^{(\mathbf{q})}(\mathbf{k}') \equiv& - V_s\Delta_s^{(\mathbf{q})}(\mathbf{k}),\\
    \sum_{\mathbf{k}'} V_{\mathbf{k},\mathbf{k}'} \mathbf{\Delta}_t^{(\mathbf{q})}(\mathbf{k}') \equiv& - V_t \mathbf{\Delta}_t^{(\mathbf{q})}(\mathbf{k})\label{eq_s15},
\end{align}
with $V_s>0$ and $V_t>0$ referring to the pair couplings in the singlet and triplet channels, respectively. The minus sign in the r.h.s. of Eqs. \eqref{eq_s14} and \eqref{eq_s15} comes from the attractive character of the interactions near the Fermi surface. Consequently, the last term in Eq. \eqref{MicAction1} becomes the following, after some algebra:
\begin{align}\nonumber
    I=&\frac{1}{2} \tr \left( \left[ \widehat{\mathscr{F}}_{\mathbf{Q}}(\mathbf{k}) \right]^\dagger \widehat{\Delta}_{\mathbf{Q}}(\mathbf{k}) \right)\\
    =& \sum_{\mathbf{p},\{\mathbf{Q}\}} \bigg( \frac{1}{V_s} |{\Delta_s^{(\mathbf{Q})}(\mathbf{p})}|^2 + \frac{1}{V_t}|{\mathbf{\Delta}_t^{(\mathbf{Q})}(\mathbf{p})}|^2\bigg).
\end{align}
We write the singlet and triplet pairing order parameters as:
\begin{align}\label{singlet-parametrization}
    \Delta_s^{(\mathbf{q})}(\mathbf{k}) =& {\Delta_s^{(\mathbf{q})}}  d_0(\mathbf{k}),\\\label{triplet-parametrization}
    \mathbf{\Delta}_t^{(\mathbf{q})}(\mathbf{k}) =& {\Delta_t^{(\mathbf{q})}}  \mathbf{d}(\mathbf{k}),
\end{align}
where ${\Delta_s^{(\mathbf{q})}}=|{\Delta_s^{(\mathbf{q})}}|e^{i\theta_s}$ and ${\Delta_t^{(\mathbf{q})}}=|{\Delta_t^{(\mathbf{q})}}|e^{i\theta_t}$, with $\theta_s$ and $\theta_t$ being, respectively, the phases of the order-parameter field in the singlet and triplet channels.
As required by Fermi-Dirac statistics, $d_0(\mathbf{k}) = d_0(-\mathbf{k})$ and $\mathbf{d}(\mathbf{k}) =-\mathbf{d}(-\mathbf{k})$. We also take the form factors $d_0(\mathbf{k})$ and $\mathbf{d}(\mathbf{k})$ to be normalized over the Fermi surface, i.e.,
\begin{align}
    \int_{FS} \frac{d \theta}{2\pi} |{d_0(\mathbf{k})}|^2=&1,\\
    \int_{FS} \frac{d \theta}{2\pi} |{\mathbf{d}(\mathbf{k})}|^2=&1.
\end{align}
Then, we obtain:
\begin{align}
    I=& \frac{1}{V_s} \sum_{\{ \mathbf{Q} \}} |{\Delta_s^{(\mathbf{Q})}}|^2+\frac{1}{V_t} \sum_{\{ \mathbf{Q}\}} |{\Delta_t^{(\mathbf{Q})}}|^2.
\end{align}
Now, we are able to write the full microscopic action as:
\begin{widetext}
\begin{align}
    \mathcal{S} =& \frac{1}{2 \beta}\sum_{\mathbf{k}, \mathbf{k}',i\omega_n} \Psi_{\mathbf{k}'}^{\dagger} \left( \left[ \widehat{\mathcal{G}}_0(\mathbf{k}, i\omega_n) \right]^{-1} \delta_{\mathbf{k},\mathbf{k}'} + \widehat{\mathcal{M}}(\mathbf{k},\mathbf{k}',\{\mathbf{Q}\}) \right) 
        \Psi_{\mathbf{k}} + \beta \sum_{\{ \mathbf{Q} \}} \left( \frac{1}{V_s} |{\Delta_s^{(\mathbf{Q})}}|^2+\frac{1}{V_t} |{\Delta_t^{(\mathbf{Q})}}|^2  \right).
\end{align}     
\end{widetext}

\section{Derivation of the Ginzburg-Landau expansion from the microscopic model}
\label{Appendix_B}

The partition function of the effective model can be calculated using the path-integral formalism. Integrating out the fermionic fields $\Psi^\dagger, \Psi$, we get (see, e.g., Ref. \cite{froldi2025}):
\begin{widetext}
\begin{align}
    \mathcal{Z}\left[ \Delta^{\dagger}, \Delta \right] = \int \mathcal{D}[\Psi^{\dagger}, \Psi] e^{-\mathcal{S}[\Psi^\dagger,\Psi]} = \det \left( \left[ \hat{\mathcal{G}}_0 \right]^{-1} + \widehat{\mathcal{M}}  \right) \exp \left[ -\frac{\beta}{2} \sum_{\{ \mathbf{Q} \}} \left( \frac{1}{V_s} |{\Delta_s^{(\mathbf{Q})}}|^2+\frac{1}{V_t} |{\Delta_t^{(\mathbf{Q})}}|^2  \right) \right].
\end{align}    
\end{widetext}

Since $\mathcal{Z}\left[ \Delta^{\dagger}, \Delta \right] = e^{-\beta \mathcal{F}\left[ \Delta^{\dagger}, \Delta \right]}$, where $\mathcal{F}\left[ \Delta^{\dagger}, \Delta \right]$ is the free energy for the order-parameter field, using $\det(A) = e^{\Tr{\ln(A)}}$, we get the Ginzburg-Landau (GL) free energy:
\begin{widetext}
    \begin{align}\label{GLFreeEnergy}
    \mathcal{F}\left[ \Delta^{\dagger}, \Delta \right] =&- \frac{1}{\beta} \Tr\{ \mathcal{G}_0^{-1} \} + \frac{1}{\beta} \sum_{n=1}^{\infty} \frac{1}{2n} \Tr{\left( \hat{\mathcal{G}}_0 \widehat{\mathcal{M}} \right)^{2n}} + \frac{1}{2}\sum_{\{ \mathbf{Q} \}} \left( \frac{1}{V_s} |{\Delta_s^{(\mathbf{Q})}}|^2+\frac{1}{V_t} |{\Delta_t^{(\mathbf{Q})}}|^2  \right),
\end{align}
\end{widetext}
where $\Tr$ stands for $\sum_{\mathbf{k},i\omega_n} \tr$, and $\tr$ refers to the trace over spin indices, and $\mathcal{F}_0 = - \frac{1}{\beta} \Tr\{ \mathcal{G}_0^{-1} \}$ is the free energy of the normal phase, which we disregard from now on. Each of the terms of order $2n$ in the sum above describes a one-particle irreducible diagram with $2n$ insertions of the complex fields {$\Delta$ or $\Delta^\dagger$}. The second-order coefficient of GL expansion describes a ``mass" term which will encode the linearized gap equation, the identification of the transition temperature, and the Cooper-pair momentum $\mathbf{Q}$ that minimizes the GL free energy. If $\mathbf{Q}=0$, one recovers a traditional zero-momentum SC phase, while if $\mathbf{Q}$ is finite, one obtains a PDW state (which can be either an FF phase or an LO phase, to be explained in more detail below). The fourth-order term of Eq. \eqref{GLFreeEnergy} describes the stability of the corresponding phase w.r.t. the interactions. In order for the GL free energy to be bounded from below, the coefficient of this fourth-order term must be positive.

Using the definitions of the order-parameter field, Eq. \eqref{FiedlDelta}, the definitions of the particle and hole propagators, Eqs. \eqref{ParticlePropagator} and \eqref{HolePropagator}, and the fact that the triplet SC states that we discuss here satisfy the unitary condition:
\begin{align}\label{UnitaryCond}
    \widehat{\mathbf{g}}^{\text{eff}}_{\mathbf{k}} \cdot \left( \boldsymbol{\Delta}^{(\mathbf{q}_i)}_t(\mathbf{k}) \times [ \boldsymbol{\Delta}^{(\mathbf{q}_i)}_t(\mathbf{k}) ]^{*} \right) = 0,
\end{align}
we calculate the second-order part of the free energy $\mathcal{F}_2$ :
\begin{widetext}
\begin{align}\label{2OGLFreeEnergy}
    \mathcal{F}_2\left[ \Delta^{\dagger}, \Delta \right] =&  \sum_{\{\mathbf{Q}\}} \left( \alpha_s(\mathbf{Q},T) |{\Delta_s^{(\mathbf{Q})}}|^2 + \alpha_t(\mathbf{Q},T) |{\Delta_t^{(\mathbf{Q})}}|^2 + \alpha_{st} (\mathbf{Q},T) |{\Delta_s^{(\mathbf{Q})}}||{\Delta_t^{(\mathbf{Q})}} |\right),
\end{align}    
where:
\begin{align}\label{alphas}
    \alpha_s (\mathbf{q},T) =& \frac{1}{V_s} +  2T\sum_{\mathbf{k},i\omega_n} d_0^2(\mathbf{k}) \left( -G^{(+)} \widetilde{G}^{(+)} + G^{(-)} \widetilde{G}^{(-)} \widehat{\mathbf{g}}^{\text{eff}}_{\mathbf{k}+\frac{\mathbf{q}}{2}} \cdot \widehat{\mathbf{g}}^{\text{eff}}_{-\mathbf{k}+\frac{\mathbf{q}}{2}} \right),\\\nonumber
    \alpha_t(\mathbf{q},T) =& \frac{1}{V_t} + 2T\sum_{\mathbf{k},i\omega_n} \bigg(  |{\mathbf{d}(\mathbf{k})}|^2 \left\{ -G^{(+)} \widetilde{G}^{(+)} - G^{(-)} \widetilde{G}^{(-)} \widehat{\mathbf{g}}^{\text{eff}}_{\mathbf{k}+\frac{\mathbf{q}}{2}} \cdot \widehat{\mathbf{g}}^{\text{eff}}_{-\mathbf{k}+\frac{\mathbf{q}}{2}} \right\}\\\label{alphat}
    &+ 2 \left[ \mathbf{d}(\mathbf{k}) \cdot \widehat{\mathbf{g}}^{\text{eff}}_{\mathbf{k}+\frac{\mathbf{q}}{2}} \right]\left[ \mathbf{d}^{*}(\mathbf{k}) \cdot \widehat{\mathbf{g}}^{\text{eff}}_{-\mathbf{k}+\frac{\mathbf{q}}{2}} \right]G^{(-)} \widetilde{G}^{(-)} \bigg),\\\nonumber
    \alpha_{st} (\mathbf{q},T) =& -4T\sum_{\mathbf{k},i\omega_n}  d_0(\mathbf{k}) \left[ \widehat{\mathbf{g}}^{\text{eff}}_{\mathbf{k}+\frac{\mathbf{q}}{2}} \times \widehat{\mathbf{g}}^{\text{eff}}_{-\mathbf{k}+ \frac{\mathbf{q}}{2}} \right] \cdot \Im {\mathbf{d}(\mathbf{k}) e^{i(\theta_s - \theta_t)}} G^{(-)} \widetilde{G}^{(-)}\\\label{S26}
    &+4 T\sum_{\mathbf{k},i\omega_n}  \Re { d_0(\mathbf{k}) \mathbf{d}(\mathbf{k}) e^{i(\theta_s - \theta_t)}} \left( G^{(+)} \widetilde{G}^{(-)} \widetilde{\mathbf{g}}^{\text{eff}}_{-\mathbf{k}+\frac{\mathbf{q}}{2}} - G^{(-)} \widetilde{G}^{(+)} \widetilde{\mathbf{g}}^{\text{eff}}_{-\mathbf{k}+\frac{\mathbf{q}}{2}} \right).
\end{align}
\end{widetext}
The linearized gap equations can be obtained from Eq. \eqref{2OGLFreeEnergy} by the minimization of the singlet and triplet order-parameter fields. Note that the term $\alpha_{st}(\mathbf{q},T)$ couples both the singlet and the triplet equations in general. However, for the cases investigated in this work, since $\widehat{\mathbf{g}}^{\text{eff}}_{\mathbf{k}} \sim \boldsymbol{\hat{z}}$, this leads to a further simplification of $\alpha_{st}(\mathbf{q},T)$,  which then displays only the second term in Eq. \eqref{S26}. Despite this, the latter term turns out to be of subleading order within the approximation that the density of states is taken to be equal to $N_F$ on both Fermi surfaces, in agreement with previous works on noncentrosymmetric superconductors (see Ref. \cite{Frigeri_2004}) and, for this reason, we neglect this term from now on. The linearized equations for the spin-singlet critical temperature $T^{(s)}_{c}$ and spin-triplet critical temperature $T^{(t)}_{c}$ are, respectively, given by $\alpha_{s}(\mathbf{Q},T_{c}^{(s)})=0$ and $\alpha_{t}(\mathbf{Q},T_{c}^{(t)})=0$, where $\mathbf{Q}$ is the previously mentioned minimizing wave vector. 

Next, we evaluate the sums and integrals of the coefficients $\alpha_s(\mathbf{q},T)$ and $\alpha_{t}(\mathbf{q},T)$ in Eqs. \eqref{alphas} and \eqref{alphat}. To this end, we follow the usual BCS procedure by considering that the interactions promoting the SC instability are in the vicinity of the Fermi surface within a range given by $-\epsilon_c< \xi_{\mathbf{k}} <\epsilon_c$. Using the fact that this cutoff is the largest energy scale of the problem, we only retain the cutoff-dependent terms and absorb them by defining the singlet and triplet critical temperatures in the absence of UPM spin splitting $\lambda$ and magnetic field $\mathbf{B}$  ({i.e.}, $\mathbf{g}^{\text{eff}}_{\mathbf{k}}=0$), respectively, by:
\begin{align}
    T^{(s)}_{c,0} =& \frac{2e^{\gamma}\epsilon_c}{\pi} e^{- 1/N_F V_s},\\
    T^{(t)}_{c,0} =& \frac{2e^{\gamma}\epsilon_c}{\pi} e^{- 1/N_F V_t},
\end{align}
where $\gamma= 0.57721...$ is the Euler-Mascheroni constant. Using the digamma function $\psi^{(0)}(x)$ to define $\Xi(x)=\Re[\psi^{(0)}\left( \frac{1}{2} + ix \right)]-\psi^{(0)}\left(\frac{1}{2} \right)$, the second-order coefficients in the GL expansion become:
\begin{widetext}
\begin{align}\nonumber
    \alpha_s(\mathbf{q},T) =& N_F\ln\left( \frac{T}{T^{(s)}_{c,0}} \right) + \frac{N_F}{4} \sum_{\gamma=\pm} \int_{FS} \frac{d\theta}{2\pi} d^2_0(\mathbf{k})\\\label{SingletAlpha}
    &\Bigg(    \Xi \left( \frac{\gamma g_1 + \frac{\mathbf{q}}{2} \cdot \mathbf{v_F}}{2 \pi T}  \right) \left( 1 + \widehat{\mathbf{g}}^{\text{eff}}_{\mathbf{k}+\frac{\mathbf{q}}{2}} \cdot \widehat{\mathbf{g}}^{\text{eff}}_{-\mathbf{k}+\frac{\mathbf{q}}{2}} \right) +  \Xi \left( \frac{\gamma g_2 + \frac{\mathbf{q}}{2} \cdot \mathbf{v_F}}{2 \pi T}  \right)  \left( 1 - \widehat{\mathbf{g}}^{\text{eff}}_{\mathbf{k}+\frac{\mathbf{q}}{2}} \cdot \widehat{\mathbf{g}}^{\text{eff}}_{-\mathbf{k}+\frac{\mathbf{q}}{2}} \right)  \Bigg),\\\nonumber
    \alpha_t(\mathbf{q},T) =& N_F \ln \left( \frac{T}{T^{(t)}_{c,0}} \right) + \frac{N_F}{4} \sum_{\gamma=\pm} \int_{FS} \frac{d\theta }{2\pi} \\\nonumber\nonumber
    &\Bigg( \abs{\mathbf{d}(\mathbf{k})}^2 \Bigg[ \Xi \left( \frac{\gamma g_1 + \frac{\mathbf{q}}{2} \cdot \mathbf{v_F}}{2 \pi T}  \right)\left( 1 - \widehat{\mathbf{g}}^{\text{eff}}_{\mathbf{k}+\frac{\mathbf{q}}{2}} \cdot \widehat{\mathbf{g}}^{\text{eff}}_{-\mathbf{k}+\frac{\mathbf{q}}{2}} \right) + \Xi \left(\frac{\gamma g_2 + \frac{\mathbf{q}}{2} \cdot \mathbf{v_F}}{2 \pi T}  \right)  \left( 1 + \widehat{\mathbf{g}}^{\text{eff}}_{\mathbf{k}+\frac{\mathbf{q}}{2}} \cdot \widehat{\mathbf{g}}^{\text{eff}}_{-\mathbf{k}+\frac{\mathbf{q}}{2}} \right) \Bigg]\\\label{TripletAlpha}
    &+ 2 \left[ \mathbf{d}(\mathbf{k}) \cdot \widehat{\mathbf{g}}^{\text{eff}}_{\mathbf{k}+\frac{\mathbf{q}}{2}} \right]\left[ \mathbf{d}(\mathbf{k}) \cdot \widehat{\mathbf{g}}^{\text{eff}}_{-\mathbf{k}+\frac{\mathbf{q}}{2}} \right] \Bigg[  \Xi \left(\frac{\gamma g_1 + \frac{\mathbf{q}}{2} \cdot \mathbf{v_F}}{2 \pi T} \right) -\Xi \left( \frac{\gamma g_2 + \frac{\mathbf{q}}{2} \cdot \mathbf{v_F}}{2 \pi T} \right) \Bigg] \Bigg).
\end{align}
\end{widetext}
In Eqs. \eqref{SingletAlpha} and \eqref{TripletAlpha}, $g_{1,2} = ( |{\mathbf{g}^{\text{eff}}_{\mathbf{k}+\mathbf{q}/2}}| \pm |{\mathbf{g}^{\text{eff}}_{-\mathbf{k}+\mathbf{q}/2}}| )/2$, and $\mathbf{v_F} = k_F(\cos\theta\,\widehat{\mathbf{x}} + \sin\theta\, \widehat{\mathbf{y}})  /m$ is the Fermi velocity at the Fermi level.

\begin{figure*}[t]
    \centering
    \includegraphics[width=0.8\linewidth]{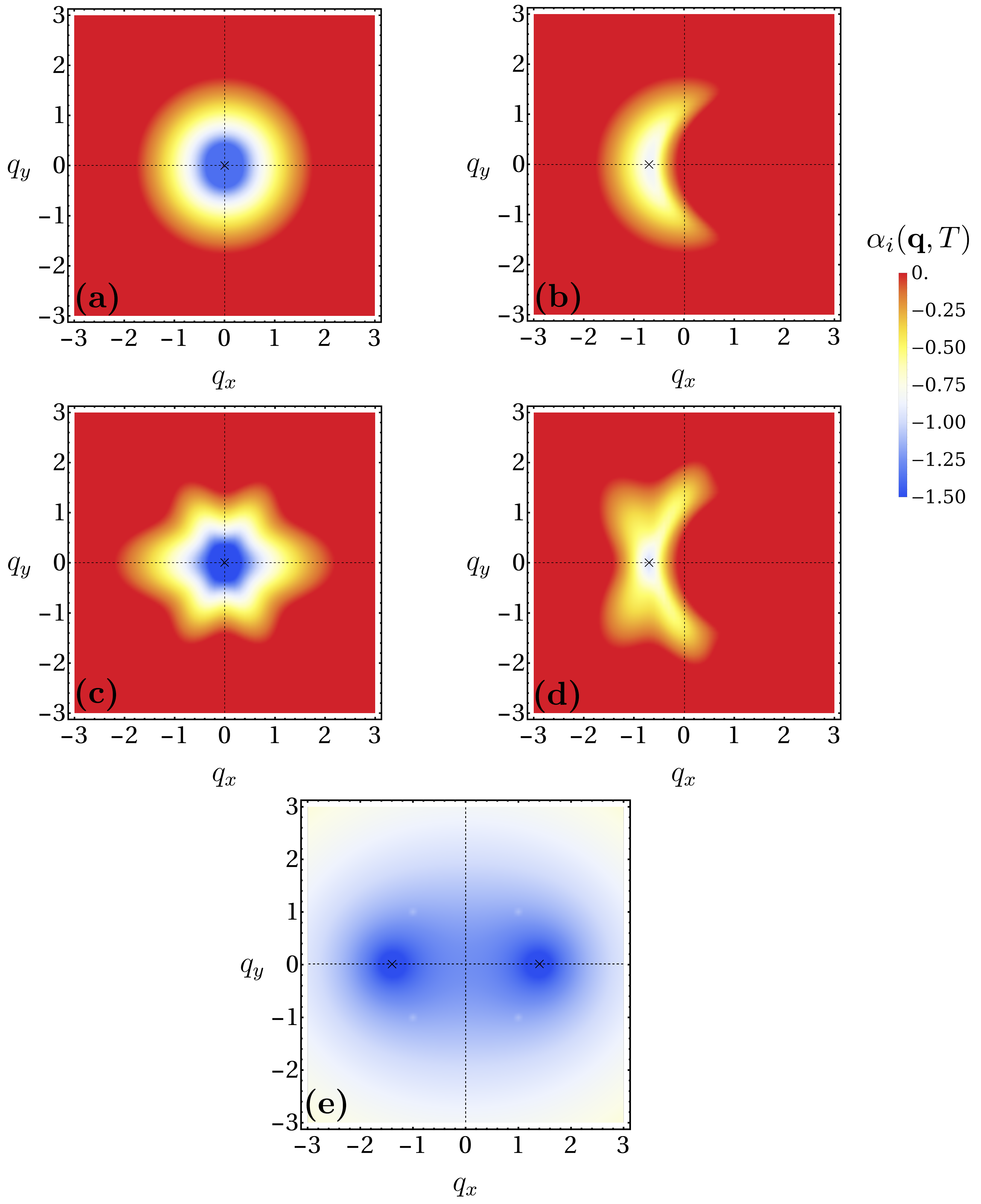}
    \caption{Second-order coefficient $\alpha_i(\mathbf{q},T)$ of the GL free energy for the singlet [{(a)}-{(d)}] (for $i=s$) and triplet [{(e)}] (for $i=t$) pairing states with fixed temperature $T=0.1T^{(i)}_{c,0}$ and splitting $\lambda/T^{(i)}_{c,0}=1$ (the minimum is represented by the symbol ``$\boldsymbol{\times}$" in those plots). Upper panels {(a)} and {(b)} are for $s$-wave states with $B_z/T^{(s)}_{c,0}=0$ and $B_z/T^{(s)}_{c,0}=1$, respectively. In the latter case, note the emergence of a single nonzero minimum in the right panel, given by $\mathbf{Q}=-0.7 \widehat{\mathbf{x}}$. Bottom panels {(c)} and {(d)} are for $d_{xy}$-wave states with $B_z/T^{(s)}_{c,0}=0$ and $B_z/T^{(s)}_{c,0}=1$, respectively. In the latter case, note the emergence of a single nonzero minimum in the right panel, given by $\mathbf{Q}=-0.7 \widehat{\mathbf{x}}$, so the only possible modulation is the FF one. Lower panel {(e)} is for helical $p$-wave state with $B_z/T^{(t)}_{c,0}=0$, showing the emergence of two minima $\pm \mathbf{Q}$, related by a $C_{2z}$ rotation.}
    \label{DensityMerged}
\end{figure*}

The plot of the coefficients \eqref{SingletAlpha} and \eqref{TripletAlpha} reveals the possible minimization wave vectors at a given temperature. For this reason, we show in Figs. \ref{DensityMerged}(a), \ref{DensityMerged}(b), \ref{DensityMerged}(c) and \ref{DensityMerged}(d), respectively, the cases for singlet $s$-wave and $d_{xy}$-wave pairing states for both zero and finite magnetic field at a fixed temperature, revealing the emergence of a single minimum in the corresponding second-order GL coefficient. In Fig. \ref{DensityMerged}(e), we show the triplet $p$-wave equal spin (i.e.,  the helical type) with gap function $\mathbf{d}(\mathbf{k})=i(k_x \boldsymbol{\hat{y}} + k_y \boldsymbol{\hat{x}})$ for zero magnetic field. The triplet $p$-wave mixed spin (i.e., the chiral type) with gap function $\mathbf{d}(\mathbf{k})=(k_x + ik_y) \boldsymbol{\hat{z}}$ yields a similar plot to those for singlet $s$-wave states.

In order to determine the critical temperature of each phase, we numerically minimize the coefficients $\alpha_s$ and $\alpha_t$ with respect to the external momentum $\mathbf{q}$, thus obtaining the temperature profile of the minimization wave vector as a function of temperature, ${\mathbf{Q}(T)}$. Then, we calculate the critical temperatures via the linearized gap equations:
\begin{align}
    \alpha_s \left( \mathbf{Q}(T^{(s)}_{c}),T^{(s)}_{c} \right) &=0,\\
    \alpha_t \left( \mathbf{Q}(T^{(t)}_{c}),T^{(t)}_{c} \right) &=0.
\end{align}

Since we have shown that our model only displays one minimization wave vector for singlet states ($\mathbf{Q}$) and at most two finite minimization wave vectors for triplet states ($\pm \mathbf{Q}$), the possible PDW phases \cite{Soto_2014,Agterberg_2020} and their spatial modulations present in the effective model investigated here are:

(a) \emph{Fulde-Ferrell} (FF) \emph{phase}: 
This pairing phase selects only one finite wave vector, thereby breaking spontaneously both time-reversal and inversion symmetries. This phase can emerge both for singlet and triplet states in the present model, and the real-space modulation of the corresponding order parameter is given by:
\begin{align}\label{DeltaFF}
{\Delta}_s^{\text{FF}}(\mathbf{r})=&\Delta^{(\mathbf{Q})}_s  e^{i\mathbf{Q}\cdot\mathbf{r}} \quad \text{(singlet FF states)},\\
{\Delta}^{\text{FF}}_t(\mathbf{r})=&{\Delta}^{(\mathbf{Q})}_t    e^{i\mathbf{Q}\cdot\mathbf{r}} \quad \text{(triplet FF states)}.
\end{align}

(b) \emph{Larkin-Ovchinnikov} {(LO)} \emph{phase}: This pairing phase is characterized by the two finite momenta, $\pm \mathbf{Q}$. Therefore, this is a time-reversal and inversion invariant version of the FF phase. This phase can only emerge for triplet states in the present model, and the real-space modulation of the  order parameter corresponding to this phase is:  
\begin{align}\label{DeltaUD}
{\Delta}^{\text{LO}}_t(\mathbf{r})= {\Delta}^{(\mathbf{Q})}_te^{i\mathbf{Q}\cdot\mathbf{r}} + {\Delta}^{(-\mathbf{Q})}_te^{-i\mathbf{Q}\cdot\mathbf{r}}  .
\end{align}

\begin{figure*}[t]
    \centering
    \includegraphics[width=0.8\linewidth]{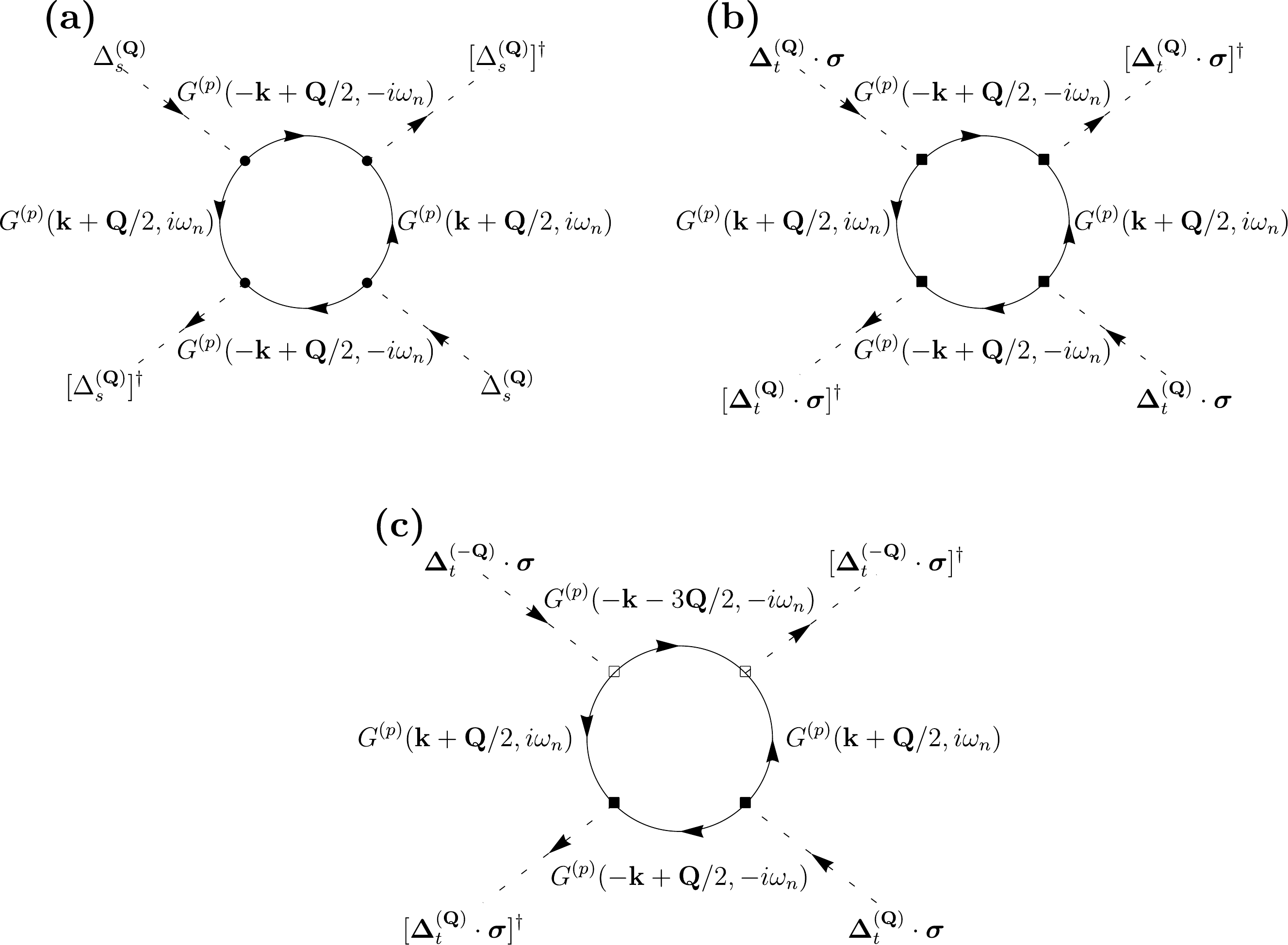}
    \caption{Fourth-order Feynman diagrams for {(a)} spin-singlet FF, {(b)} spin-triplet FF, and {(c)} spin-triplet LO order. The Feynman rules for the vertices are extracted from Eq. \eqref{GeneralFourthOrder} as $\bullet \equiv d_0(\mathbf{k}) \sigma_y$ for incoming external legs, $\bullet \equiv [d_0(\mathbf{k}) \sigma_y]^{\dagger}$ for outgoing external legs, $\blacksquare \equiv \mathbf{d}(\mathbf{k}) \cdot \boldsymbol{\sigma} \sigma_y$ for incoming external legs, $\blacksquare \equiv [\mathbf{d}(\mathbf{k}) \cdot \boldsymbol{\sigma} \sigma_y]^\dagger$ for outgoing external legs, $\square \equiv \mathbf{d}(\mathbf{k}+\mathbf{Q}) \cdot \boldsymbol{\sigma} \sigma_y$ for incoming external legs, and $\square \equiv [\mathbf{d}(\mathbf{k}+\mathbf{Q}) \cdot \boldsymbol{\sigma} \sigma_y]^\dagger$ for outgoing external legs. Since this is a pairing channel, the following similarity transformation $G^{(h)}(\mathbf{k},i\omega_n) = - [G^{(p)}(-\mathbf{k},-i\omega_n)]^{T}$ was used to derive these diagrams. }
    \label{FF-UD-Feynman}
\end{figure*}

We now include the fourth-order term of the GL free energy, which is given by:
\begin{widetext}
    \begin{align}\nonumber
    \delta \mathcal{F}^{(4)}=& \frac{T}{2} \sum_{\mathbf{k}, i\omega_n} \sideset{}{'}\sum_{\substack{\mathbf{q_1},\mathbf{q_2}\\\mathbf{q_3} }} \Tr \Bigg\{ G_{0}^{(p)} \left( \mathbf{k} \right) \widehat{\Delta}^{(\mathbf{q}_1)} \left(\mathbf{k} - \frac{\mathbf{q}_1}{2} \right)  \sigma_y G_{0}^{(h)} \left( \mathbf{k} - \mathbf{q}_1 \right) \left[ \widehat{\Delta}^{(\mathbf{q}_2)} \left(\mathbf{k}-\mathbf{q}_1 + \frac{\mathbf{q}_2}{2}\right)   \sigma_y \right] ^{\dagger} G_{0}^{(p)} \left( \mathbf{k}- \mathbf{q}_1+ \mathbf{q}_2 \right) \\\label{GeneralFourthOrder}
    &  \widehat{\Delta}^{(\mathbf{q}_3)} \left( \mathbf{k} - \mathbf{q}_1 + \mathbf{q}_2 - \frac{\mathbf{q}_3}{2} \right)   \sigma_y G_{0}^{(h)} \left( \mathbf{k} -\mathbf{q}_1 + \mathbf{q}_2 - \mathbf{q}_3 \right) \left[ \widehat{\Delta}^{(\mathbf{q}_1 - \mathbf{q}_2 + \mathbf{q}_3)} \left( \mathbf{k} - \frac{\mathbf{q}_1}{2} + \frac{\mathbf{q}_2}{2} - \frac{\mathbf{q}_3}{2} \right)   \sigma_y \right]^{\dagger} \Bigg\},
\end{align}
\end{widetext}
where the prime in the summation above means that there is an additional constraint given by $\mathbf{q}_1 +\mathbf{q}_2 - \mathbf{q}_3 \in \mathcal{A}$, where $\mathcal{A}=\{\pm\mathbf{Q}\}$ is the set of minimization wave vectors. The calculation of these expressions can be summarized in terms of  Feynman diagrams, which are shown in Fig. \ref{FF-UD-Feynman} for both FF (singlet and triplet) and LO (triplet) phases. We will calculate these fourth-order coefficients below for the spin-singlet and spin-triplet pairing states, separately. 

\section{Fourth-order coefficient of the GL expansion for singlet pairing states}
\label{Appendix_C}

We first focus on spin-singlet pairing states by making $\mathbf{d}(\mathbf{k})=0$. For completeness, we consider both $s$-wave and $d_{xy}$-wave pairing symmetries (for wave vectors in the vicinity of the $\Gamma$ point), i.e., with form factors given by  
\begin{align}
    d_0(\mathbf{k}) =& 1 \Rightarrow s\text{-wave},\\
    d_0(\mathbf{k}) =& \frac{2\sqrt{2}}{k^2_F}k_x k_y \Rightarrow d_{xy}\text{-wave}.
\end{align}
Consequently, we observe that only an FF phase can emerge upon the application of a magnetic field in these cases. Therefore, the only contribution that comes from the fourth-order coefficient is obtained by setting $\mathbf{q}_i = \mathbf{Q}$ (for $i=1,2,3$) in Eq. \eqref{GeneralFourthOrder}, or, equivalently, evaluating the Feynman diagram in Fig. \ref{FF-UD-Feynman}(a), which yields:
\begin{widetext}
\begin{align}\nonumber
    \delta \mathcal{F}_{FF,singlet}^{(4)}=& |{\Delta_s^{(\mathbf{Q})}}|^{4} T \sum_{\mathbf{k}, i\omega_n} d_0^{4}(\mathbf{k}) \Bigg\{ \left[ G^{(+)} \widetilde{G}^{(+)} \right]^{2} - 2G^{(+)} \widetilde{G}^{(+)} G^{(-)} \widetilde{G}^{(-)} \widehat{\mathbf{g}}^{\text{eff}} \cdot \widetilde{\mathbf{g}}^{\text{eff}} +\left[ G^{(+)} \widetilde{G}^{(-)} \right]^2\\\nonumber
    & - 2G^{(+)} \widetilde{G}^{(-)} G^{(-)} \widetilde{G}^{(+)} \widehat{\mathbf{g}}^{\text{eff}} \cdot \widetilde{\mathbf{g}}^{\text{eff}}  + \left[ G^{(-)} \widetilde{G}^{(+)} \right]^{2}\\
    &+ \left[ G^{(-)} \widetilde{G}^{(-)} \right]^{2}\left[  \left( \widehat{\mathbf{g}}^{\text{eff}} \cdot \widetilde{\mathbf{g}}^{\text{eff}} \right)^2 - (\widehat{\mathbf{g}}^{\text{eff}} \times \widetilde{\mathbf{g}}^{\text{eff}}) \cdot (\widehat{\mathbf{g}}^{\text{eff}} \times \widetilde{\mathbf{g}}^{\text{eff}}) \right] \Bigg\}\equiv  \chi_{s} (\mathbf{Q}) |{\Delta_s}^{(\mathbf{Q})}|^{4},
\end{align}
\end{widetext}
where we have used the notation $\widehat{\mathbf{g}}^{\text{eff}} \equiv \widehat{\mathbf{g}}^{\text{eff}}_{\mathbf{k}+\frac{\mathbf{q}}{2}}$ and $\widetilde{\mathbf{g}}^{\text{eff}} \equiv \widehat{\mathbf{g}}^{\text{eff}}_{-\mathbf{k}+\frac{\mathbf{q}}{2}}$. The pairing phase diagram for the case of spin-singlet $s$-wave pairing is shown in the main text in the limit of $J_{sd}\ll \lambda$. Regarding the phase diagram for the case of spin-singlet $d_{xy}$-wave pairing, its structure is similar to the pairing phase diagram for $s$-wave singlet states, with only a slightly smaller critical temperature compared to the latter.


\begin{figure}[t]
    \centering
    \includegraphics[width=1\linewidth]{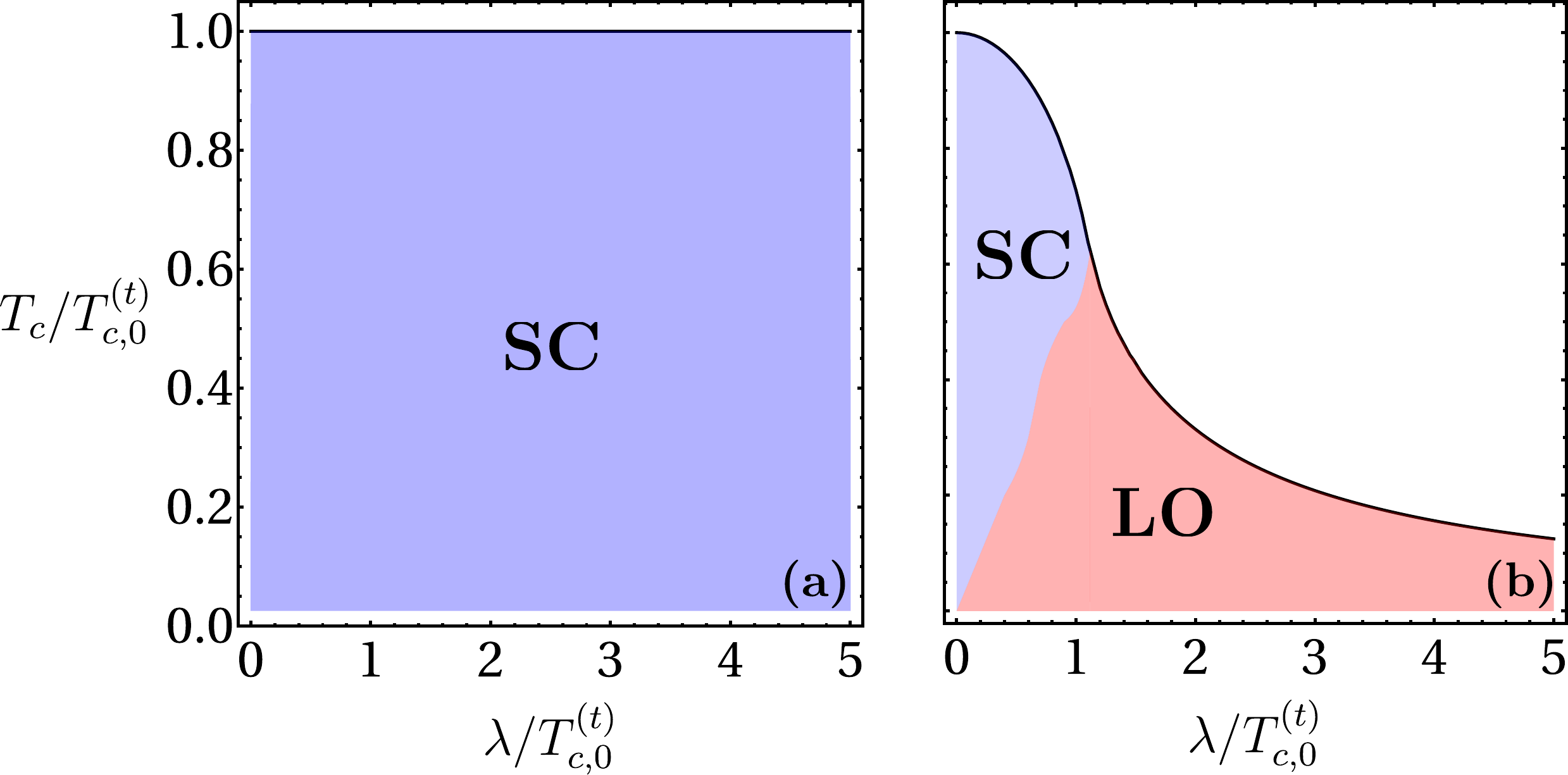}
    \caption{Triplet critical temperatures of the chiral $p$-wave (a) and helical $p$-wave (b) pairing states as a function of the UPM splitting $\lambda$ for zero magnetic field. Note that the chiral $p$-wave state is not affected by $\lambda$ and also does not generate any finite-momentum superconductivity. An LO state only emerges in the case of a helical $p$-wave phase for finite $\lambda$. Although the latter state is suppressed by $\lambda$, this suppression is mild, since $T_c$ only reaches zero in the limit $\lambda \rightarrow\infty$.}
    \label{TcTripletZeroB}
\end{figure}

\section{Fourth-order coefficient of the GL expansion for triplet pairing states}
\label{Appendix_D}

We now focus on the triplet states by making $d_0(\mathbf{k})=0$. As explained in the main text, we consider both helical $p$-wave and chiral $p$-wave pairing symmetries (for wave vectors in the vicinity of the $\Gamma$ point), i.e., with form factors given by:
\begin{align}
    \mathbf{d}(\mathbf{k}) =& i(k_x \boldsymbol{\hat{y}} +k_y \boldsymbol{\hat{x}}) \Rightarrow \text{helical }p\text{-wave},\\
    \mathbf{d}(\mathbf{k}) =& (k_x + ik_y) \boldsymbol{\hat{z}} \Rightarrow \text{chiral }p\text{-wave}.
\end{align}
Note from Fig. \ref{DensityMerged}(e) that we can have either one or two possible minimization wave vectors $\pm \mathbf{Q}$, so we need to evaluate the fourth-order coefficient for triplet states given in the Feynman diagrams shown in Figs. \ref{FF-UD-Feynman}(b) and \ref{FF-UD-Feynman}(c). Evaluating these diagrams, we get for the FF phase:
\begin{widetext}
\begin{align}\nonumber
    \delta \mathcal{F}_{FF,triplet}^{(4)} &= |{\Delta_{t}^{(\mathbf{Q})}}|^{4} T \sum_{\mathbf{k},\omega_n} \Bigg( [G^{(+)}\widetilde{G}^{(+)}]^2[2(\mathbf{d} \cdot \mathbf{d}^*)^2 - (\mathbf{d} \cdot \mathbf{d})(\mathbf{d}^* \cdot \mathbf{d}^*)] + 2[G^{(-)}\widetilde{G}^{(+)}]^2[ (\mathbf{d} \cdot \mathbf{d})(\mathbf{d}^* \cdot \mathbf{d}^*)] \\\nonumber
    &+ [G^{(-)}\tilde{G}^{(-)}]^2[2((\widehat{\mathbf{g}}^{\text{eff}}\cdot\mathbf{d})(\widetilde{\mathbf{g}}^{\text{eff}}\cdot\mathbf{d}^*) - (\widehat{\mathbf{g}}\times\mathbf{d})\cdot(\widetilde{\mathbf{g}}^{\text{eff}}\times\mathbf{d}^*))^2 - (\mathbf{d} \cdot \mathbf{d})(\mathbf{d}^* \cdot \mathbf{d}^*)] \\\nonumber
    &- 4G^{(+)}G^{(-)}\widetilde{G}^{(+)}\widetilde{G}^{(-)}(\mathbf{d} \cdot \mathbf{d}^*)(\widehat{\mathbf{g}}^{\text{eff}}\cdot\mathbf{d})(\widetilde{\mathbf{g}}^{\text{eff}}\cdot\mathbf{d}^*) - (\widehat{\mathbf{g}}^{\text{eff}}\times\mathbf{d})\cdot(\widetilde{\mathbf{g}}^{\text{eff}}\times\mathbf{d}^*)]  \Bigg)\\
    &\equiv  \chi^{\text{FF}}_{t} (\mathbf{Q}) |{\Delta_{t}^{(\mathbf{Q})}}|^{4},
\end{align}    
\end{widetext}
where $\mathbf{d} \equiv \mathbf{d}(\mathbf{k})$. For the LO phase, two copies of the diagram in Fig. \ref{FF-UD-Feynman}(b) appear, one corresponding to $+\mathbf{Q}$ and the other to $-\mathbf{Q}$, and other four expressions appear corresponding to the diagram in Fig. \ref{FF-UD-Feynman}(c). Using the corresponding Feynman rules, we obtain:
\begin{widetext}
\begin{align}\nonumber
   \delta \mathcal{F}_{\text{LO},triplet}^{(4)} &= \chi^{\text{FF}}_{t} (\mathbf{Q}) |{\Delta_{t}^{(\mathbf{Q})}}|^{4} + \chi^{\text{FF}}_{t} (-\mathbf{Q}) |{\Delta_{t}^{(-\mathbf{Q})}}|^{4}\\\nonumber
   & + 4 |{\Delta_{t}^{(\mathbf{Q})}}|^{2} |{\Delta_{t}^{(-\mathbf{Q})}}|^{2} T \sum_{\mathbf{k},\omega_n} \tr \Bigg\{ \left( G^{(+)} \sigma_0 + G^{(-)} \widehat{\mathbf{g}}^{\text{eff}}_{\mathbf{k}+\frac{\mathbf{Q}}{2}} \cdot \boldsymbol{\sigma} \right) \mathbf{d}(\mathbf{k}) \cdot \boldsymbol{\sigma} \sigma_y \\\nonumber
   &\left( \widetilde{G}^{(+)} \sigma_0 + \widetilde{G}^{(-)}\widehat{\mathbf{g}}^{\text{eff}}_{-\mathbf{k}+\frac{\mathbf{Q}}{2}} \cdot \boldsymbol{\sigma} \right) [\mathbf{d}(\mathbf{k}) \cdot \boldsymbol{\sigma} \sigma_y]^{\dagger}  \left( G^{(+)} \sigma_0 + G^{(-)} \widehat{\mathbf{g}}^{\text{eff}}_{\mathbf{k}+\frac{\mathbf{Q}}{2}} \cdot \boldsymbol{\sigma} \right) \mathbf{d}(\mathbf{k}+\mathbf{Q}) \cdot \boldsymbol{\sigma} \sigma_y \\\nonumber
   &\left( G'^{(+)} \sigma_0 + G'^{(-)} \widehat{\mathbf{g}}^{\text{eff}}_{-\mathbf{k}-\frac{3\mathbf{Q}}{2}} \cdot \boldsymbol{\sigma} \right) [\mathbf{d}(\mathbf{k}+\mathbf{Q}) \cdot \boldsymbol{\sigma} \sigma_y]^{\dagger}\Bigg\}\\
   &= \chi^{\text{FF}}_{t} (\mathbf{Q}) |{\Delta_{t}^{(\mathbf{Q})}}|^{4} + \chi^{\text{FF}}_{t} (-\mathbf{Q}) |{\Delta_{t}^{(-\mathbf{Q})}}|^{4} + 4 |{\Delta_{t}^{(\mathbf{Q})}}|^{2} |{\Delta_{t}^{(-\mathbf{Q})}}|^{2}\chi_{t}^{\text{LO}} (\mathbf{Q}),
\end{align}
\end{widetext}
where we have used the notation $G^{(\pm)}(\mathbf{k}+\frac{\mathbf{q}}{2}) \equiv G^{(\pm)}$, $G^{(\pm)}(-\mathbf{k}+\frac{\mathbf{q}}{2}) \equiv \widetilde{G}^{(\pm)}$ and $G'^{(+/-)} \equiv G'^{(+/-)}\left( -\mathbf{k}-\frac{3\mathbf{Q}}{2}, -i\omega_n \right)$.

\begin{figure*}[t]
    \centering
    \includegraphics[width=1\linewidth]{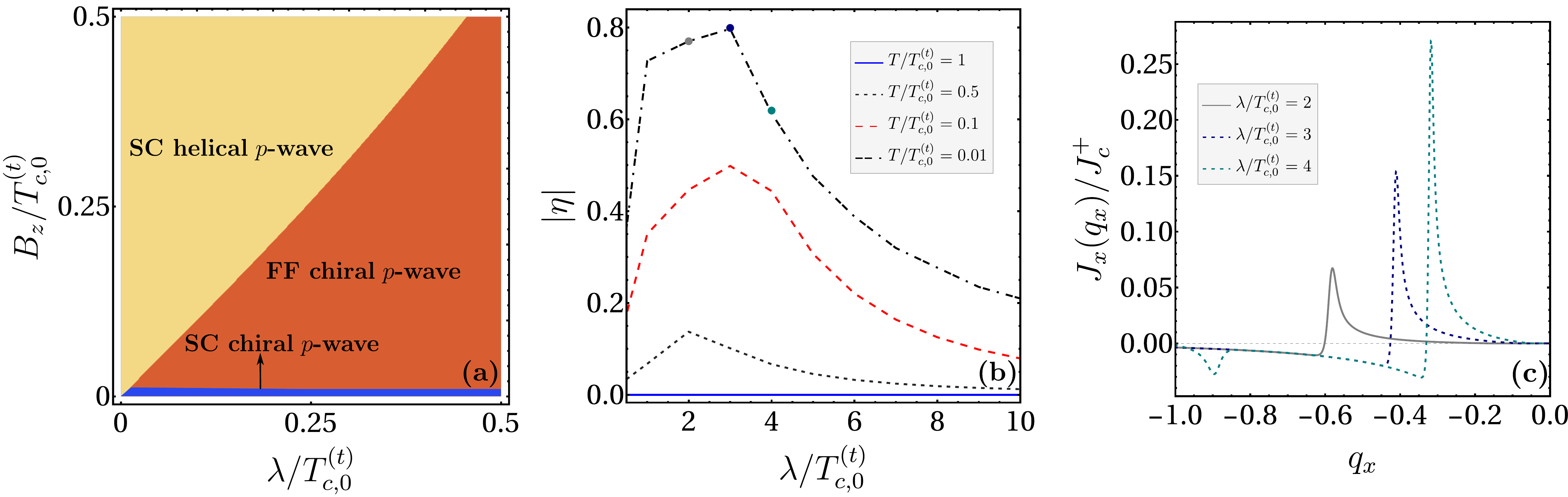}
    \caption{{(a)} Schematic pairing phase diagram of $B_z$ vs $\lambda$ in the present model for spin-triplet superconducting states. {(b)} Efficiency $\eta $ of the SDE for the chiral $p$-wave FF phase with modulation along the direction of the minimizing wave vector $\mathbf{Q}$ as a function of the UPM spin splitting $\lambda$ for a finite magnetic field such that $ B_z/T^{(t)}_{c,0} =1$. We also show the behavior of $\eta$ in the model for different temperatures $T$. {(c)} Supercurrent as a function of the momentum $q_x$ for different values of the spin splitting around the maximum of the efficiency, indicated by three circles shown in (b), at  $T=0.01T^{(t)}_{c,0}$.}
    \label{EffTriplet}
\end{figure*}

We are now able to determine whether the emergent PDW phase in Fig. \ref{DensityMerged}(e) is an FF or an LO phase, by checking which has the smaller free energy. The analysis of energetics within the GL theory reveals that for the helical $p$-wave state, the stable phase is the LO phase, as shown in Fig. \ref{TcTripletZeroB}(b). Note, however, that the chiral $p$-wave phase is unaffected by the UPM spin splitting, whereas the helical $p$-wave state is mildly suppressed. The corresponding triplet pairing phase diagram is displayed in Fig. \ref{EffTriplet}(a), where we have obtained which pairing phase has the highest critical temperature, revealing the dominant order. As a result, the LO pairing phase is not present in the phase diagram in Fig. \ref{EffTriplet}(a). 

\section{Superconducting diode efficiencies of other pairing symmetries}
\label{Appendix_E}

\begin{figure}[t]
    \centering
    \includegraphics[width=0.75\linewidth]{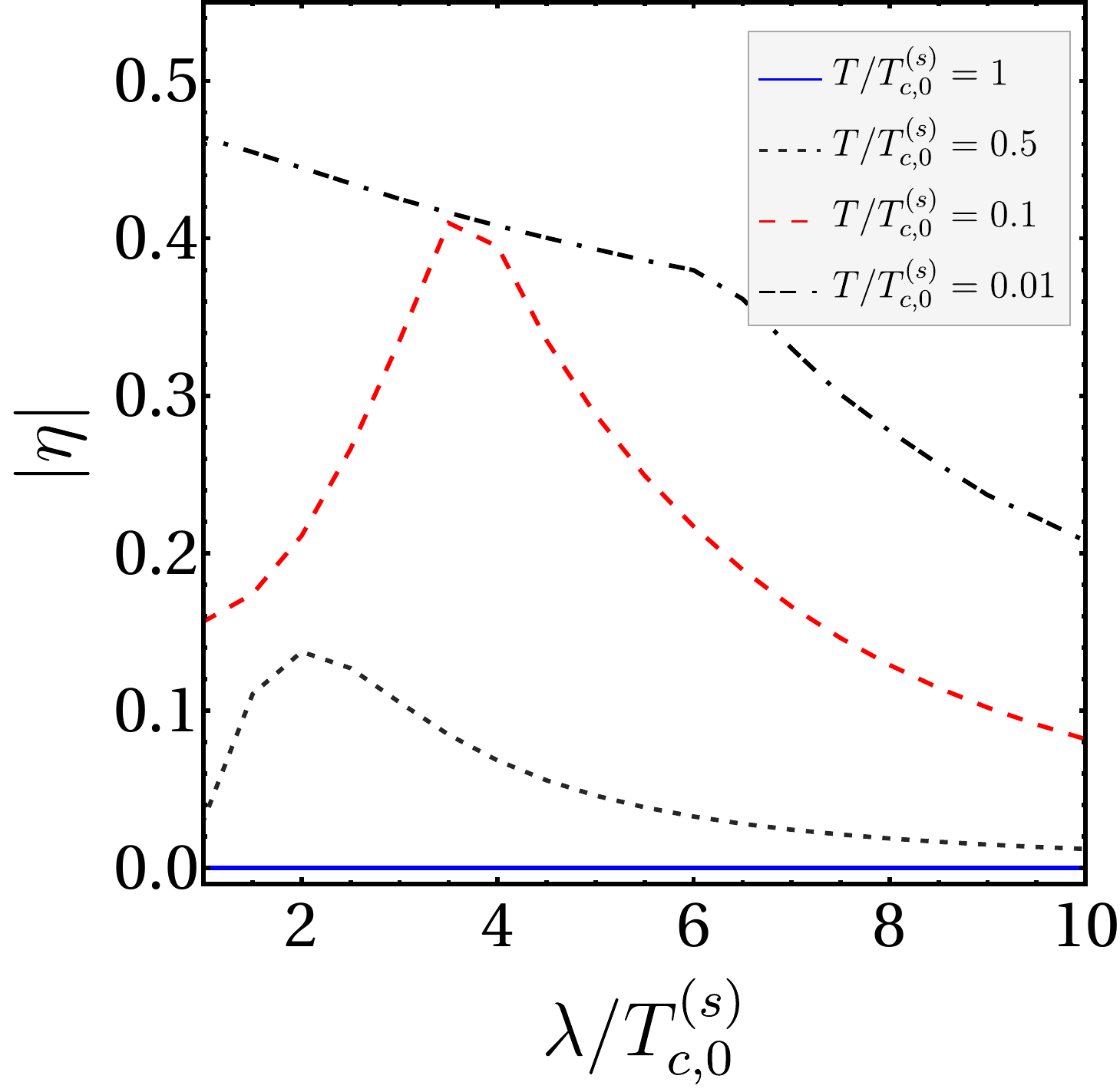}
    \caption{Efficiency $\eta $ of the SDE for the spin-singlet $d_{xy}$-wave FF phase with modulation along the direction of the minimizing wave vector $\mathbf{Q}$ as a function of the UPM spin splitting $\lambda$ for a finite magnetic field such that $ B_z/T^{(s)}_{c,0} =1$. We also show the behavior of $\eta$ in the model for different temperatures $T$.}
    \label{EffTriplet2}
\end{figure}

We turn our attention to the SDE efficiencies in our model for the additional pairing symmetries investigated in this work. As explained in the main text, only the FF state will lead to a finite intrinsic diode efficiency $\eta$, since this pairing phase breaks both time-reversal and inversion symmetries. For this reason, in Fig. \ref{EffTriplet2}, we show the corresponding efficiency for the spin-singlet $d_{xy}$-wave FF state in the presence of a finite magnetic field. As a result, we observe that the SDE efficiency in this case yields  smaller values than the $s$-wave FF case, which was discussed in the main text. 

\begin{figure*}[t]
    \centering
    \includegraphics[width=0.69\linewidth]{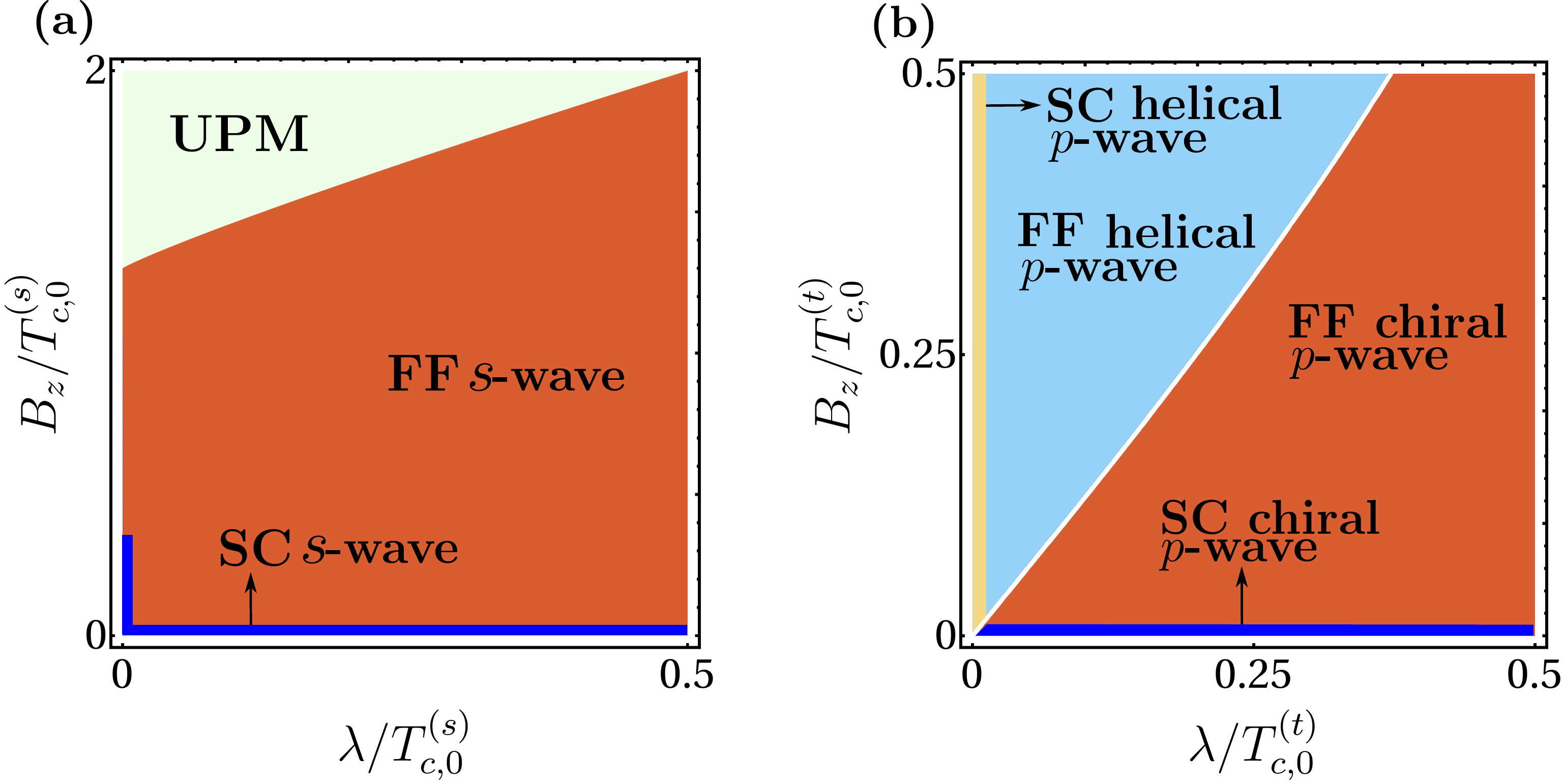}
    \caption{(a) Schematic pairing phase diagram of $B_z$ vs $\lambda$ in the present model for $J_{sd}/T^{(s)}_{c,0}=1$. The white region denotes the UPM phase, the red region stands for the spin-singlet $s$-wave FF phase, and the dark blue region corresponds to the spin-singlet $s$-wave zero-momentum SC phase. (b) Schematic pairing phase diagram of $B_z$ vs $\lambda$ in the present model for spin-triplet $p$-wave superconducting states for $J_{sd}/T^{(t)}_{c,0}=1$. The labels explain the corresponding leading superconducting phases.}
    \label{PhaseDiagramJsdSinglet}
\end{figure*}

In Fig. \ref{EffTriplet}(b), we show the corresponding efficiency for the chiral $p$-wave FF state. Note that this quantity is quantitatively similar to the efficiencies obtained for the spin-singlet $s$-wave FF state explained in the main text. Indeed, we observe here that the SDE efficiency also has a peak for low temperatures at an optimal splitting, which can reach almost $80\%$ of efficiency for some conveniently chosen physical parameters. We also point out that the SDE efficiency in this case  exhibits a tendency to increase as the temperature decreases. Finally, in Fig. \ref{EffTriplet}(c), the supercurrents are plotted at $T=0.01T^{(t)}_{c,0}$ as a function of the component of the wave vector $q_x$, which refers to the direction of modulation $\mathbf{Q}$ that produces the highest SDE efficiencies.

\begin{figure}[t]
    \centering
    \includegraphics[width=0.85\linewidth]{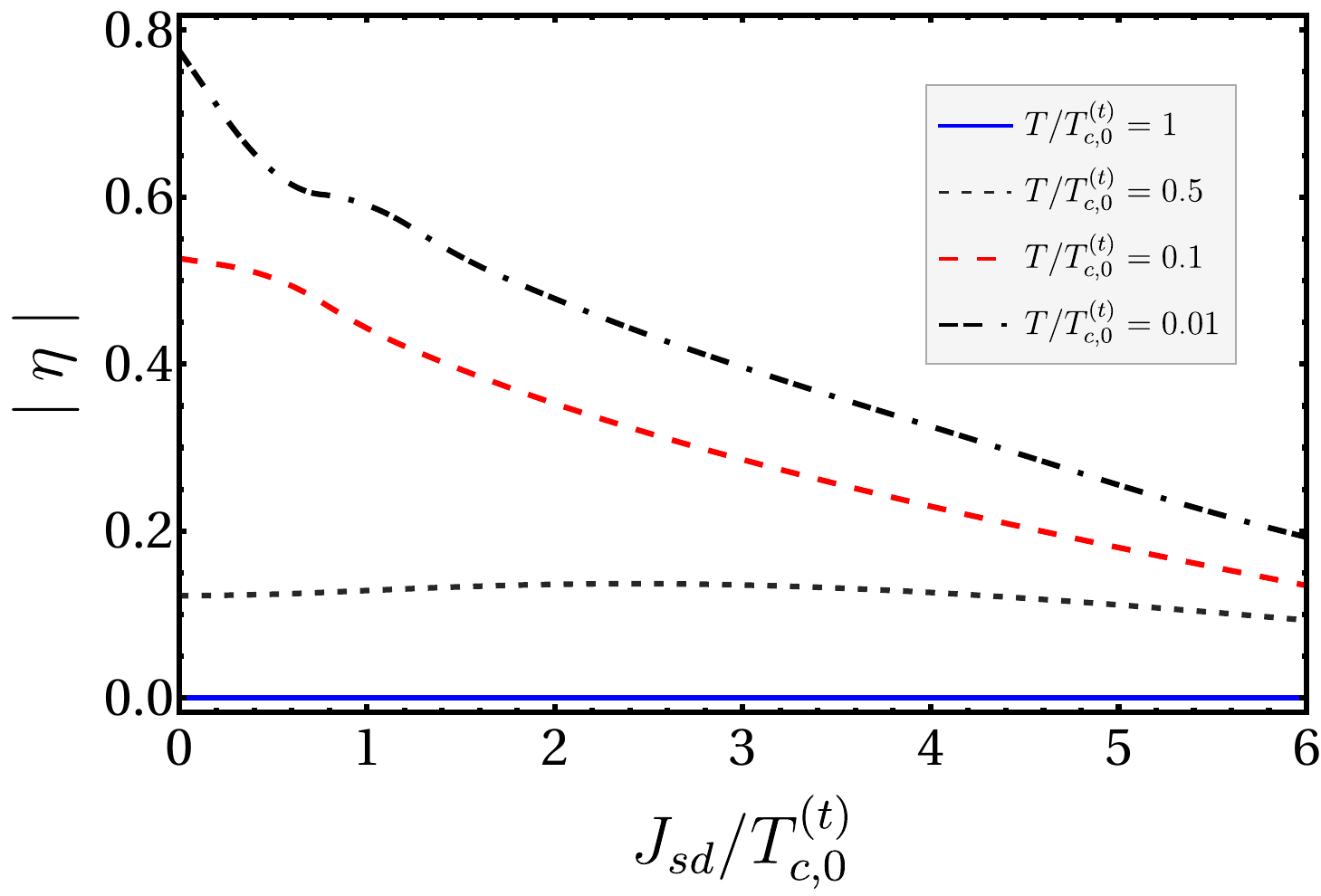}
    \caption{SDE efficiencies as a function of the $J_{sd}$ coupling in Eq. \eqref{HUPM} associated with different values of the temperature for the chiral $p$-wave FF phase with modulation along the direction of the minimizing wave vector $\mathbf{Q}$, a finite magnetic field $B_z/T_{c,0}^{(t)}=1$, and spin splitting $\lambda/T_{c,0}^{(t)}=3$.}
    \label{EffJsdTriplet}
\end{figure}

\section{Effect of a moderate $J_{sd}$ coupling in the model}
\label{Appendix_F}

In this last section, we discuss the effect of a moderate $J_{sd}$ in the effective model defined in Eq. \eqref{HUPM}. In Fig. \ref{PhaseDiagramJsdSinglet}, we show the corresponding pairing phase diagrams for both $s$-wave spin-singlet and $p$-wave spin-triplet superconducting phases. As can be seen from the plot in the case of spin-singlet pairing for a moderate $J_{sd}/T^{(s)}_{c,0}=1$ [Fig. \ref{PhaseDiagramJsdSinglet}(a)], the phase diagram does not change significantly from the diagram shown in the main text in the limit of $J_{sd}\ll \lambda$. The main differences are a mild suppression of the SC phase for $\lambda=0$ and finite $B_z$, and a decrease of the critical out-of-plane magnetic field that drives the transition from an FF-phase to a normal UPM metal. The SDE efficiency corresponding to this value of $J_{sd}$ is also shown in the main text.

By contrast, for the case of $p$-wave spin-triplet pairing, the inclusion of a moderate $J_{sd}$ in the effective model defined in Eq. \eqref{HUPM} leads to some qualitative changes in the corresponding phase diagram. Indeed, by comparing Fig. \ref{PhaseDiagramJsdSinglet}(b) to Fig. \ref{EffTriplet}(a), we note that the main difference between these two diagrams consists  in the emergence of a helical $p$-wave SC phase only for $\lambda=0$ and finite $B_z$, and the appearance of a new helical $p$-wave FF state for small $\lambda$ and large $B_z$. As for the remaining triplet pairing phases, the chiral $p$-wave SC phase is essentially unchanged for a moderate $J_{sd}$ coupling in the model, and the region of the phase diagram displaying the chiral $p$-wave FF state becomes larger. Finally, we show in Fig. \ref{EffJsdTriplet} the effect of the $J_{sd}$ coupling in the SDE efficiency of the chiral $p$-wave FF phase for different temperatures.


%

\end{document}